\documentclass[11pt,letterpaper]{article}

\usepackage[T1]{fontenc}
\usepackage{textcomp}
\usepackage[centertags]{amsmath}
\usepackage{amsfonts}
\usepackage{amsthm}
\usepackage{amssymb}
\usepackage[numbers,sort&compress]{natbib}
\usepackage[hypertex]{hyperref}
\usepackage{calc}

\usepackage{paperinitial}
\paperinitialization{15mm}{15mm}{20mm}{10mm}{2pt}{10pt}

\DeclareMathOperator{\pr}{pr}

\DeclareMathOperator{\Sp}{Sp}
\newcommand{\lan}{\langle}
\newcommand{\ran}{\rangle}

\newcommand{\e}{\varepsilon}
\newcommand{\vf}{\varphi}
\newcommand{\vk}{\varkappa}
\newcommand{\s}{\sigma}

\newcommand{\Si}{\Sigma}
\newcommand{\al}{\alpha}
\newcommand{\be}{\beta}

\newcommand{\Ga}{\Gamma}
\newcommand{\de}{\delta}
\newcommand{\De}{\Delta}

\newcommand{\la}{\lambda}
\newcommand{\La}{\Lambda}
\newcommand{\ups}{\upsilon}

\newcommand{\spx}{\mathbf{x}}
\newcommand{\spy}{\mathbf{y}}
\newcommand{\spp}{\mathbf{p}}
\newcommand{\spk}{\mathbf{k}}

\newcommand{\ddo}{\dot{0}}

\begin{document}
\setlength{\unitlength}{1pt}
\allowdisplaybreaks[4]

\title{{\Large\textbf{Ultraviolet asymptotics of particle creation\\with respect to a congruence of observers}}}

\date{}

\author{P.O. Kazinski\thanks{E-mail: \texttt{kpo@phys.tsu.ru}}\\[0.5em]
{\normalsize Physics Faculty, Tomsk State University, Tomsk 634050, Russia}}

\maketitle

\begin{abstract}

The construction of quantum field theory (QFT) of a free massive scalar field with respect to a general congruence of observers is considered, the splitting into positive- and negative-frequency modes being defined by diagonalization of the instantaneous Hamiltonian. The explicit expression for the ultraviolet asymptotics of the average number of particles created from the vacuum is found. It is shown that, for a general congruence of observers in the $D$-dimensional spacetime with $D\geqslant3$, the total number of created particles diverges in the ultraviolet domain in the regularization removal limit. This holds even in the Minkowski spacetime. Therefore, in this case, the quantum evolution is not unitary in the regularization removal limit. It is proved that not all classically admissible congruences of observers are proper on a quantum level. Namely, unitarity of QFT with respect to a congruence of observers is violated when the equal time hypersurfaces of this congruence are not spacelike, even for a theory with a finite cutoff. The implications of these results are discussed.


\end{abstract}




\section{Introduction}

The issues with unitarity of a free quantum evolution of massive fields in globally hyperbolic spacetimes have a long history. It was recognized already in \cite{Park1} that, for certain splitting of modes of a quantum field into positive- and negative-frequency ones, the total number of particles created from a vacuum can be infinite for non-stationary cosmological spacetimes such as FLRW. This infinity does not stem from a poor infrared behavior of a theory but arises in the ultraviolet domain for both massive and massless particles. In physical terms, it appears that the average number of particles produced from the vacuum by a nonstationary metric field declines too slowly, as $\omega^{-2}$, at large energies $\omega$ \cite{Park1,FullingAQFT,GriMaMos.11,ParkTom,Parkrev}. In the relatively recent paper \cite{TorVara}, it was shown that such divergencies arise even in the Minkowski spacetime for certain unfortunate choices of the splitting of the scalar field into positive- and negative-frequency parts. So that unitarity of evolution is violated in the $D$-dimensional spacetime with $D\geqslant3$ in this case. In the present paper, we construct quantum field theory (QFT) of a scalar field with respect to a general congruence of observers with the creation-annihilation operators defined by diagonalization of the instantaneous Hamiltonian (this gives the representation of quantum fields in the Fock space) and arrive at the same conclusion. Namely, for a general congruence of observers, the average number of particles created from the vacuum behaves as $\omega^{-2}$ in the ultraviolet spectral domain even in the Minkowski spacetime, and so unitarity is violated in the regularization removal limit for $D\geqslant3$. This clearly shows that QFT strongly depends on a choice of a congruence of observers. As for quantum theories with a finite cutoff, there are no such problems with unitarity. However, in this case, the Heisenberg equations depend explicitly on the cutoff and the choice of a congruence of observers and do not reduce to the Klein-Gordon equation in the ultraviolet domain.

The issues with unitarity described above spring from the two sources. First, it is the assumption that the modes with \emph{arbitrary} high energies, higher, for example, than the Planck scale, propagate according to the Klein-Gordon equation or its analog for higher spins. Second, it is the definition of creation-annihilation operators or, in other words, the choice of the representation of quantum fields in the Fock space. In the overwhelming majority of papers and books, the first point is taken for granted, which, in particular, is expressed in the use of quantum fields in the Heisenberg representation obeying exactly the Klein-Gordon equation. This assumption also underlies the motivation for introducing the Hadamard states (see, e.g., \cite{KayWald}). As far the second point is concerned, there was developed a procedure to define the so-called adiabatic vacuum that allows one to avoid the unitarity problem \cite{Park1,FullingAQFT,ParkTom,Parkrev,BirDav.11,CalzHu}. We shall describe briefly this procedure and stress its drawbacks that are relevant for our consideration.

First of all, note that the unitary problem can formally be solved if one defines the splitting of the modes in the following way (see, e.g., \cite{Scharf79,AgAsht}). Let we have, at a given instant of time, some splitting of the modes into positive- and negative-frequency ones. Then we evolve these modes into the future using the Klein-Gordon equation. Call a mode at the instant of time $t$ a positive(negative)-frequency one if it results from a positive(negative)-frequency mode at the initial instant of time. Then, by definition, the particle creation is absent and unitarity holds. The $n$th order adiabatic creation-annihilation operators are associated with the modes constructed as above but with the replacement of the exact mode functions by their $n$th order (in the inverse energy) WKB approximation. Taking $n$ sufficiently large, one can secure unitarity. The above definitions of creation-annihilation operators both the easiest and the adiabatic ones, although formally perfect, possesses obvious drawbacks. They are nonlocal in time, i.e., the definition of the Fock space, where the operators act, is not determined by the state of the background fields at the present moment. One has to know the whole evolution since the beginning of time up to the present moment. Furthermore, different choices of the ``beginning of time'' give rise to unitary inequivalent quantum theories, in general. This is a rather strange feature. Another drawback is that the adiabatic definition is ambiguous as it depends on the order parameter $n$, and the expansion in increasing $n$ is asymptotic \cite{ParkTom}.

As opposed to the common wisdom, in the present paper we explicitly introduce the cutoff into the Hamiltonian, since it is evident that the modes with the energies higher than the Planck scale are not described reliably by the Klein-Gordon equation. This solves the problem with unitarity unless the cutoff is not removed. Besides, we use the method of diagonalization of the instantaneous Hamiltonian to define the creation-annihilation operators. This method was introduced in \cite{GriMaMos.11,Imamur,Shirok0,Shirok,GrMa} and criticized in \cite{Park1,FullingAQFT,junker}. Why do we stick to this method despite the fact that it was blamed in Refs. \cite{Park1,FullingAQFT,ParkTom,BirDav.11,CalzHu,junker}? The main reason is that the ground state of the instantaneous Hamiltonian, i.e., the vacuum of the corresponding Fock space, can be prepared with an arbitrary high precision. This fact follows from the standard adiabatic theorems of quantum mechanics (see, e.g., \cite{LandLifshQM.11,Kato} and for recent results \cite{JoyeThes,Nenciu,AveElg,HagJoye,JoyeGener,ElgHag}) under rather general assumptions that evolution is unitary and its time-dependent generators $\hat{H}(t)$ have a common dense domain. In particular, if the system starts from the vacuum state of $\hat{H}(0)$, then the adiabatic theorems guarantee that the distance between the actual state of the system at the time $t$ and the ground state of the instantaneous Hamiltonian $\hat{H}(t)$ is of the order $O(\tau^{-1})$, where $\tau$ is a characteristic time of variations of the background fields. For the adiabatic theorems can be applied, $\tau$ should be much larger than the inverse energy gap between the instantaneous vacuum and the first excited state. This gap equals approximately to the mass $m$ of a particle, and so we have a fairly small quantity $m^{-1}\approx 1.3\times 10^{-21}$ s for the vacuum of electrons and positrons. The common dense domain of $\hat{H}(t)$ and unitarity of quantum evolution can always be achieved by introducing a cutoff. It looks more plausible that the equation governing quantum dynamics changes above the Planck scale rather than the postulates of quantum theory are not valid, i.e., unitarity is violated or the Hamiltonian is not defined for physically realizable systems. The second reason is that such definition of the representation in the Fock space is local in time, i.e., it is determined by the state of the background fields at a given instant of time. For a particular case of a static metric with a static congruence of observers, this method reproduces the standard definition of creation-annihilation operators. Of course, as soon as the creation-annihilation operators and the representation of the fields in the Fock space are defined, one may perform arbitrary unitary transformations in the respective Fock space to go to other sets of creation-annihilation operators.

Notice also that the issues with unitarity we are discussing are peculiar only to the gravitational interaction and absent, even in the regularization removal limit, for the electromagnetic background with the definition of particles given by the instantaneous Hamiltonian diagonalization \cite{Scharf79}. As we shall see, this is a consequence of the fact that the strength of the gravitational interaction on a tree level grows linearly with the energy of a particle (the equivalence principle, see, e.g., \cite{WeinbIR}), while, in the electromagnetic case, it does not depend on the energy. Mathematically, the metric field enters the principal symbol of the Klein-Gordon operator, in contrast to the electromagnetic field, that results in a more intense particle creation by the gravitational field in the ultraviolet spectral range.

Another important point we concern in the present paper is the explicit inclusion of observers into a theory. In spite of the fact that it is well recognized \cite{Zelman.12,DeWQG1,Vladimir.12,Mitskev.12,KucharI,KucharII,Isham.12,KucharPoT} that a congruence of observers is a necessary ingredient of general relativity to provide measurability of the metric, the curvature, the energy-momentum tensor, and so on, often these observers are cast out from a theory. This is rather harmless on a classical level, though one has to introduce them implicitly or explicitly to obtain observable quantities from general relativity. On the quantum side, as we shall see, it is crucial to introduce the congruence of observers from the outset, since different congruences result in unitary inequivalent theories in the regularization removal limit. In a theory with a cutoff, the dependence of observables on a choice of the congruence of observers is evident. Therefore, in both cases, we have to take into account the presence of a congruence of observers in the theory.

In Sec. \ref{Hamilt_Diag}, we start with the construction of quantum theory of a boson field with a cutoff on a nonstationary background by means of diagonalization of the instantaneous Hamiltonian. For the most part, the formalism presented in Sec. \ref{Hamilt_Diag} was already given in \cite{KazMil1,KazMil2}. We, however, reproduce these formulas in the present paper to make it self-contained and to correct some mistakes made in \cite{KazMil1,KazMil2}. In the next Sec. \ref{Scal_Field}, we apply the general formalism to a massive scalar field evolving with respect to a general congruence of observers. In fact, we consider the scalar field in the Minkowski spacetime. Nevertheless, it turns out that the main conclusions do not depend on the curvature of the metric. We also prove in this section that classically proper congruences of observers with non-spacelike hypersurfaces of an equal time cannot by used to construct QFT, because such a theory is inevitably non-unitary even with a finite cutoff. Section \ref{Scal_Field} is concluded by the general formula for the average number of particles $n_\al$ created from a vacuum. It is written in terms of the commutator Green function and the eigenvectors associated with the instantaneous Hamiltonian (the mode functions). In Sec. \ref{ULtr_Asympt}, we derive the ultraviolet WKB asymptotic expressions for these mode functions. Then the ultraviolet asymptotics of $n_\al$ is estimated for a general congruence of observers. It is shown in this case that $n_\al\sim\omega^{-2}$ and so the total number of created particles diverges for $D\geqslant3$. Notice that the powers of $\omega$ of the asymptotic expansion at large energies and the coefficients of this expansion follow rigorously from the standard theorems of the WKB method (see, e.g., \cite{Fedoryuk.6,BabBuld,MaslovCWKB,BBT.1,BBTZ}). At the end of this section, we prove that the creation-annihilation operators associated with the instantaneous Hamiltonian taken at different times cannot be realized in one Fock space. Therefore, the quantum evolution takes place not in one Fock space but in the Hilbert bundle, its fibers being the Fock spaces and the base representing the infinite dimensional space of background fields. This picture has to be accepted even in quantum electrodynamics with respect to inertial observers in the Minkowski spacetime. Otherwise, the evolution of a quantum Dirac field in an external nonstationary magnetic field is not unitary \cite{Ruijsen}. The way that allows one to reduce this evolution to one Fock space, if needed, is discussed in Sec. \ref{Hamilt_Diag}. Section \ref{Pert_Anal} is devoted to a perturbative approach to the problem. Namely, it is assumed that the congruence of observers differ slightly from a congruence of inertial observers in the Minkowski spacetime. In that case, the general formulas of Sec. \ref{ULtr_Asympt} can be simplified and the explicit expression for the leading ultraviolet asymptotics of $n_\al$ is obtained. The peculiarity (non-perturbativity) of the gravitational interaction in the ultraviolet domain that was mentioned above is clearly seen at this stage. In Conclusion, we discuss some implication of the obtained results. These results are especially important for the background field method in spacetimes with Lorentzian signature, since the knowledge of free dynamics of quantum fields on a given classical background can be used to obtain the one-loop effective action \cite{BirDav.11,GriMaMos.11,BuchOdinShap.11,WeinB2,DeWGAQFT.11,GFSh.3,GrMuRaf}.

\section{General formulas}\label{Hamilt_Diag}

In this section, we briefly recall the main steps of the construction of QFT by means of the Hamiltonian diagonalization procedure. The idea to use the Hamiltonian diagonalization in order to define the notion of a particle was proposed in \cite{GriMaMos.11,Imamur,Shirok0,Shirok,GrMa}. We shall follow the procedure elaborated in  \cite{KazMil1,KazMil2} correcting some mistakes made in these papers. Our approach is close to the one developed in \cite{Shirok} for a quantum Dirac field interacting with an external electromagnetic field.

Let the Hamiltonian of a scalar quantum field in the Schr\"{o}dinger representation be
\begin{equation}\label{Hamilt_quadr}
    \hat{H}(t)=\frac12 \hat{Z}^AH_{AB}(t)\hat{Z}^B,\qquad \hat{Z}^A=\left[
                                           \begin{array}{c}
                                             \hat{\phi}(\spx) \\
                                             \hat{\pi}(\spx) \\
                                           \end{array}
                                         \right],\qquad
    [\hat{Z}^A,\hat{Z}^B]=iJ^{AB}=\left[
                        \begin{array}{cc}
                          0 & i \\
                          -i & 0 \\
                        \end{array}
                      \right]\de(\spx-\spy).
\end{equation}
We assume that
\begin{equation}
    \bar{H}_{AB}=H_{BA}=H_{AB},
\end{equation}
The bar over the expression means a complex conjugation. Usually, $H_{AB}$ defines a positive-definite quadratic form. Nevertheless, we do not imply this property.

Let us pose the eigenvalue problem for the non-singular self-conjugate operator $-iJ_{AB}=(iJ^{AB})^{-1}$ with respect to the quadratic form $H_{AB}$:
\begin{equation}\label{eigen_prblm}
    -iJ_{AB}\ups^B_\al(t)=\omega_\al^{-1}(t) H_{AB}(t)\ups^B_\al(t),\qquad \omega_\al^{-1}(t) \bar{\ups}^A_\al(t) H_{AB}(t)\ups^B_\al(t)>0,
\end{equation}
where $\ups_\al^A$ obey certain boundary conditions following from the problem statement. Inasmuch as $J_{AB}$ is non-singular, $\omega_\al^{-1}\neq0$. Moreover, we assume that the spectrum is real and discrete, and, for any $\La>0$, there exits a finite number of eigenvalues such that $\omega_\al<\La$. The inequality in \eqref{eigen_prblm} is the definition of the splitting of the modes into positive-frequency $\ups_\al$ and negative-frequency $\bar{\ups}_\al$ ones. Notice that $\omega_\al(t)$ can be negative when $H_{AB}(t)$ is not positive-definite.

The orthogonality and completeness relations read as
\begin{equation}\label{orth_complt}
    \{\ups_\al,\ups_\be\}=\{\bar{\ups}_\al,\bar{\ups}_\be\}=0,\quad\{\ups_\al,\bar{\ups}_\be\}=-i\de_{\al\be},\qquad iJ^{AB}=\sum_\al\ups_\al^{[A}\bar{\ups}_\al^{B]},
\end{equation}
where $\{\ups,w\}:=J_{AB}\ups^A w^B$. The square brackets at a pair of indices means antisymmetrization without the one-half factor. The normalization of eigenfunctions is chosen to be
\begin{equation}
    \bar{\ups}^A_\al H_{AB}\ups_\al^B=\ups^A_\al H_{AB}\bar{\ups}_\al^B=\omega_\al.
\end{equation}
In other words, the vectors ($\ups_\al$,$\bar{\ups}_\al$) constitute a symplectic basis. In the case when $H_{AB}(t)$ is positive-definite, the above properties of the spectrum and the mode functions are met with $\omega_\al(t)>0$ provided the background fields are sufficiently smooth and the problem is posed in the domain of the variables $\spx$ of a finite volume defined by the use of the Euclidean metric $\de_{ij}$ (see more precise definitions in Secs. \ref{Scal_Field}, \ref{ULtr_Asympt}). Here we just assume that these properties of the mode functions are valid.

Then we introduce the creation-annihilation operators
\begin{equation}\label{creaannh_oper}
    \hat{a}_\al(t):=\{\bar{\ups}_\al(t),\hat{Z}\},\qquad \hat{a}^\dag_\al(t):=\{\ups_\al(t),\hat{Z}\}.
\end{equation}
The completeness relation \eqref{orth_complt} implies
\begin{equation}\label{ZinAAd}
    \hat{Z}^A=-i\sum_\al(\ups_\al^A\hat{a}_\al-\bar{\ups}_\al^A\hat{a}^\dag_\al).
\end{equation}
Substituting \eqref{ZinAAd} into \eqref{Hamilt_quadr}, we find
\begin{equation}\label{Hamilton_diagonalized}
    \hat{H}=\frac12\sum_\al\omega_\al(t)[\hat{a}^\dag_\al(t)\hat{a}_\al(t)+\hat{a}_\al(t) \hat{a}^\dag_\al(t)].
\end{equation}
As a rule, this operator is not defined in the Fock space $F_t$ with the vacuum annihilated by $\hat{a}_\al(t)$. Therefore, we define the regularized Hamilton operator
\begin{equation}\label{Hamilt_reg}
    \hat{H}_\La(t)=\frac12\sum_\al^\La\omega_\al(t)[\hat{a}^\dag_\al(t)\hat{a}_\al(t)+\hat{a}_\al(t) \hat{a}^\dag_\al(t)]=\sum_\al^\La \omega_\al(t)\hat{a}^\dag_\al(t)\hat{a}_\al(t)+\frac12\sum_\al^\La\omega_\al(t),
\end{equation}
where the sum over $\al$ is finite and is carried over those eigenvalues that correspond to the energies $\omega_\al(t)<\La$. Let us stress that the Hilbert space where the regularized Hamiltonian \eqref{Hamilt_reg} acts is infinite dimensional and is not truncated, but the creation-annihilation operators corresponding to the high-energy modes do not enter into \eqref{Hamilt_reg}. Of course, the sharp cutoff is by no means a distinguished one. One can replace $\theta(\La-\omega_\al(t))\omega_\al(t)$ in \eqref{Hamilt_reg} by $f(\omega_\al(t)/\La)\omega_\al(t)$, where $f(x)$ is a smooth positive cutoff function such that $f(x)\approx 1$ for $x\ll1$ and $f(x)$ tends sufficiently fast to zero for $x\gg1$.

The creation-annihilation operators corresponding to the different instants of time $t$ and $t_{in}$ are related by the canonical transform
\begin{equation}
    \left[
      \begin{array}{c}
        \hat{a}(t) \\
        \hat{a}^\dag(t) \\
      \end{array}
    \right]=
    \left[
      \begin{array}{cc}
        F(t,t_{in}) & G(t,t_{in}) \\
        \bar{G}(t,t_{in}) & \bar{F}(t,t_{in}) \\
      \end{array}
    \right]
    \left[
      \begin{array}{c}
        \hat{a}(t_{in}) \\
        \hat{a}^\dag(t_{in}) \\
      \end{array}
    \right],
\end{equation}
where
\begin{equation}\label{FandG}
    F_{\al\be}=-i\{\bar{\ups}_\al(t),\ups_\be(t_{in})\},\qquad G_{\al\be}=i\{\bar{\ups}_\al(t),\bar{\ups}_\be(t_{in})\}.
\end{equation}
This canonical transform is a unitary one if and only if $G_{\al\be}$ is the Hilbert-Schmidt operator \cite{Shale,BerezMSQ1.4}, i.e.,
\begin{equation}
    \Sp G^\dag G<\infty.
\end{equation}
In Sec. \ref{ULtr_Asympt}, we shall show that for a scalar field on the time-dependent metric background of a general form the operator $G$ is not Hilbert-Schmidt. Therefore, $\hat{a}_\al(t)$ act in different Fock spaces for the different $t$.

In the case we are interested in, the mode functions and, hence, the creation-annihilation operators depend on $t$ only through the background fields entering into the instantaneous Hamiltonian \eqref{Hamilt_quadr}. Therefore, we introduce the Hilbert bundle, with the base being the space of background fields and the fibers $F_t$ representing the Fock spaces\footnote{The other Hilbert bundle was introduced in \cite{AgAsht,AsCaPa} to describe the quantum evolution. It is the bundle over a real line representing the time variable.}. In this bundle, we define the unitary parallel transport operator $\hat{W}_{t,t_{in}}:F_{t_{in}}\rightarrow F_{t}$ (cf. \cite{Kato}) such that
\begin{equation}\label{parall_transp_def}
    \hat{a}_\al(t)=\hat{W}_{t,t_{in}}\hat{a}_\al(t_{in})\hat{W}_{t_{in},t},\qquad |vac,t\ran:=\hat{W}_{t,t_{in}}|vac,t_{in}\ran,
\end{equation}
where $|vac,t\ran$ is the vacuum vector in $F_t$. This operator is unitary and obeys the equation
\begin{multline}\label{parall_transp}
    -i\partial_t\hat{W}_{t,t_{in}}=\frac12\sum_{\al,\be}\Big[2\hat{a}^\dag_\al(t)\{\dot{\bar{\ups}}_\al(t),\ups_\be(t)\} \hat{a}_\be(t)\\
    -\hat{a}_\al(t)\{\dot{\ups}_\al(t),\ups_\be(t)\}\hat{a}_\be(t) -\hat{a}^\dag_\al(t)\{\dot{\bar{\ups}}_\al(t),\bar{\ups}_\be(t)\}\hat{a}_\be^\dag(t)\Big]\hat{W}_{t,t_{in}},
\end{multline}
with the initial condition $\hat{W}_{t_{in},t_{in}}=1$. Notice that the sums over $\al$ and $\be$ in \eqref{parall_transp} are not truncated by a cutoff. Using the relations \eqref{orth_complt}, it is easy to verify that the connection
\begin{equation}\label{connection}
    \hat{\Ga}_\mu:=\frac12\sum_{\al,\be}\Big[2\hat{a}^\dag_\al \Big\{\frac{\de\bar{\ups}_\al}{\de\Phi^\mu},\ups_\be \Big\} \hat{a}_\be
-\hat{a}_\al \Big\{\frac{\de\ups_\al}{\de\Phi^\mu },\ups_\be \Big\}\hat{a}_\be -\hat{a}^\dag_\al \Big\{\frac{\de\bar{\ups}_\al}{\de\Phi^\mu},\bar{\ups}_\be \Big\}\hat{a}_\be^\dag\Big]
\end{equation}
is a trivial one. Here $\Phi^\mu$ is a shorthand notation for all the background fields. Therefore, if the fundamental group of the space of background fields is trivial (this is a rather mild assumption as such space is infinite dimensional), then the parallel transport $\hat{W}_{t,t_{in}}\equiv \hat{W}_{\Phi(t),\Phi(t_{in})}$ does not depend on the way how the points $\Phi(t)$ and $\Phi(t_{in})$ are connected. The important property of the definition of particles by means of diagonalization of the Hamiltonian is the locality in time, i.e., the Fock space $F_t$, where the measurements at a given instant of time are performed, is determined only by the value of the background fields at this instant of time. This property is a necessary condition for the Schr\"{o}dinger representation to be properly defined. In the general case, when all the quantum fields acquire their own classical background counterparts, $\hat{\Phi}^\mu\rightarrow\Phi^\mu+\de\hat{\Phi}^\mu$,
one can think of the quantum evolution as a curve in the tangent bundle of the phase space of the background fields. Every fiber of this bundle is equipped with the structure of a Fock space.

The evolution operator $\hat{U}^\La_{t,t_{in}}$ maps the Fock space $F_{t_{in}}$ into $F_{t}$. We can bring this evolution into one Fock space with the aid of the unitary parallel transport \eqref{parall_transp_def}. The physically measurable amplitudes are the matrix elements of the operator
\begin{equation}\label{Smatr_energy_def}
    \hat{S}^\La_{t,t_{in}}:=\hat{W}_{t_{in},t}\hat{U}^\La_{t,t_{in}}
\end{equation}
in the Fock space $F_{t_{in}}$. This operator satisfies the equation
\begin{multline}\label{Smatr_energy_1}
    i\partial_t\hat{S}^\La_{t,t_{in}}=\Big\{\frac12\sum_\al^\La\omega_\al(t) [\hat{a}^\dag_\al(t_{in})\hat{a}_\al(t_{in})+\hat{a}_\al(t_{in}) \hat{a}^\dag_\al(t_{in})]\\ +\sum_{\al,\be}\big[\hat{a}^\dag_\al(t_{in})\{\dot{\bar{\ups}}_\al,\ups_\be\}\hat{a}_\be(t_{in}) -\frac12\hat{a}_\al(t_{in})\{\dot{\ups}_\al,\ups_\be\}\hat{a}_\be(t_{in}) -\frac12\hat{a}^\dag_\al(t_{in})\{\dot{\bar{\ups}}_\al,\bar{\ups}_\be\}\hat{a}_\be^\dag(t_{in})\big] \Big\}\hat{S}^\La_{t,t_{in}}
\end{multline}
with the initial condition $\hat{S}^\La_{t_{in},t_{in}}=1$.

The parallel transport operator is determined by \eqref{parall_transp_def} only up to a phase, and equation \eqref{parall_transp} corresponds to a particular choice of it. The operator
\begin{equation}\label{add_phases}
    \hat{W}'_{t,t_{in}}=e^{i\al[\Phi(t)]-i\al[\Phi(t_{in})]}\hat{W}_{t,t_{in}},
\end{equation}
where $\al[\Phi]$ is a real-valued functional, is quite as good as $\hat{W}_{t,t_{in}}$. That leads to the ambiguity in the definition of the amplitudes \eqref{Smatr_energy_def}. However, this ambiguity is unimportant since it cannot be observed in experiments. The transition probability does not depend on this phase. As for the phase difference that can be observed in interferometric-like experiments, in this case the system is driven along the two paths $\Phi_1(t)$ and $\Phi_2(t)$ with the same initial and final points. Therefore, $\al[\Phi]$ does not contribute to the phase difference either.

The results of the paper \cite{Scharf79} generalized to a scalar field imply that $\hat{S}^\La_{t,t_{in}}$ is a unitary operator for sufficiently good background electromagnetic fields in the regularization removal limit, the counterterms renormalizing the contributions to the vacuum polarization being assumed to be added to \eqref{Hamilt_reg}. In order to employ these results, we need to pass to the Heisenberg representation (see below). As regards a scalar field on a time-dependent metric background, the  $\hat{S}^\La_{t,t_{in}}$ is not unitary, in general, for both finite and infinite cutoffs.

The evolution $\hat{S}^\La_{t,t_{in}}$ can always be made unitary by introducing the energy cutoff into the sums over $\al$ and $\be$ in the second line of \eqref{Smatr_energy_1}. Let us denote this operator as $\hat{S}'^\La_{t,t_{in}}$. It is clear that the mode functions corresponding to the energies above, say, the Planck scale are of a little physical importance since for those energies the very Hamiltonian \eqref{Hamilt_quadr} becomes invalid. Then the parallel transport operator $\hat{W}_{t,t_{in}}$ can be replaced by its regularized version $\hat{W}^\La_{t,t_{in}}$ obeying
\begin{equation}\label{parall_transp_reg}
\begin{split}
    i\partial_t\hat{W}^\La_{t_{in},t}=&\frac12\sum_{\al,\be}\Big[\hat{a}^\dag_\al(t_{in})\big(\{\dot{\bar{r}}_\al(t),r_\be(t)\} -\{\bar{r}_\al(t),\dot{r}_\be(t)\}\big) \hat{a}_\be(t_{in})-\\
    &-\hat{a}_\al(t_{in})\{\dot{r}_\al(t),r_\be(t)\}\hat{a}_\be(t_{in}) -\hat{a}^\dag_\al(t_{in})\{\dot{\bar{r}}_\al(t),\bar{r}_\be(t)\}\hat{a}_\be^\dag(t_{in})\Big]\hat{W}^\La_{t_{in},t},
\end{split}
\end{equation}
where $\hat{W}^\La_{t,t}=1$ and
\begin{equation}\label{u_al}
    r_\al(t):=\theta(\La-\omega_\al(t))\ups_\al(t).
\end{equation}
The delta-functions arising from the differentiation of the theta-functions are canceled due to the relations \eqref{orth_complt}, and the sums over $\al$ and $\be$ in \eqref{parall_transp_reg} are finite. If there is a finite number of modes with $\omega_\al(t)<\La$ for some $t$ during the whole evolution, the problems \eqref{Smatr_energy_1}, \eqref{parall_transp_reg} correspond, in essence, to a system with a finite number degrees of freedom and define unitary operators in one and the same Fock space $F_{t_{in}}$ (all the Fock spaces connected by $\hat{W}^\La_{t,t_{in}}$ become unitary equivalent). As the mode functions \eqref{u_al} do not satisfy the completeness relation \eqref{orth_complt}, the corresponding connection $\hat{\Ga}^\La_\mu$ possesses a non-zero curvature, but this curvature tends to zero in the regularization removal limit. When such a regularization is introduced, the evolution operator $\hat{U}^\La_{t,t_{in}}:=\hat{W}^\La_{t,t_{in}} \hat{S}'^\La_{t,t_{in}}$ maps $F_{t_{in}}$ into $F_{t_{in}}$ and is unitary. Then, as we have already discussed in Introduction, the standard adiabatic theorems can be applied to this evolution (see, e.g., \cite{LandLifshQM.11,Kato,JoyeThes,Nenciu,AveElg,HagJoye,JoyeGener,ElgHag}). If the system evolves from the vacuum state, the uniform adiabatic theorem says that the distance in the Hilbert space between the instantaneous vacuum and the actual state of the system is of the order $O(\tau^{-1})$, where $\tau$ is a characteristic time scale of the background field variations. This time scale must be much larger than the inverse gap between the vacuum state and the first excited one, which of the order $m^{-1}$. For example, for the vacuum state of electrons and positrons and more massive particles $m^{-1}<1.3\times 10^{-21}$ s. Taking $\tau\gg m^{-1}$, one can prepare the system in the state very close to the instantaneous vacuum. This procedure shows that the instantaneous vacua \eqref{parall_transp_def} and their excitations possess a real physical meaning.

Further, we shall employ a more elegant construction based on the geometric picture described above. Namely, we define the evolution operator $\hat{U}^\La_{t,t_{in}}:F_{t_{in}}\rightarrow F_t$ as
\begin{equation}\label{evolut_reg}
    \hat{U}^\La_{t,t_{in}}:=\hat{W}_{t,t_{in}} \hat{S}'^\La_{t,t_{in}}.
\end{equation}
By construction $\hat{S}'^\La_{t,t_{in}}$ and $\hat{W}_{t,t_{in}}$ are unitary. The regularized evolution \eqref{evolut_reg} satisfies the Schr\"{o}dinger equation
\begin{equation}
    i\partial_t\hat{U}^\La_{t,t_{in}}=\hat{H}'_\La(t)\hat{U}^\La_{t,t_{in}}
\end{equation}
with the generator
\begin{multline}\label{Hamilt_reg1}
    \hat{H}'_\La(t)=\frac12\sum_\al^\La\omega_\al(t)[\hat{a}^\dag_\al(t)\hat{a}_\al(t)+\hat{a}_\al(t) \hat{a}^\dag_\al(t)]-\\
    -\sideset{}{'}\sum_{\al,\be}\big[\hat{a}^\dag_\al(t)\{\dot{\bar{\ups}}_\al,\ups_\be\}\hat{a}_\be(t) -\frac12\hat{a}_\al(t)\{\dot{\ups}_\al,\ups_\be\}\hat{a}_\be(t) -\frac12\hat{a}^\dag_\al(t)\{\dot{\bar{\ups}}_\al,\bar{\ups}_\be\}\hat{a}_\be^\dag(t)\big],
\end{multline}
where the prime at the sum sign reminds us that only such $\al$ and $\be$ are left in the sum that either $\omega_\al(t)$ or $\omega_\be(t)$ or both of them are larger than $\La$. The term on the second line in \eqref{Hamilt_reg1} can be considered as a counterterm. It is self-adjoint, local in time, and disappears in the regularization removal limit. Its inclusion makes $\hat{S}'^\La_{t,t_{in}}$ unitary in the ultraviolet domain and provides the adiabatic evolution for the high-energy modes. The regularized Hamiltonian \eqref{Hamilt_reg1} can be written in terms of the field operators $\hat{Z}^A$, if one substitutes \eqref{creaannh_oper} into \eqref{Hamilt_reg1}.

Now we pass into a more familiar Heisenberg representation
\begin{equation}\label{CAO_Heis}
\begin{gathered}
    \hat{a}_\al(in)=\hat{a}_\al(t_{in}),\qquad
    \hat{a}_\al(out)=\hat{U}^\La_{t_{in},t_{out}}\hat{a}_\al(t_{out})\hat{U}^\La_{t_{out},t_{in}} =\hat{S}'^\La_{t_{in},t_{out}}\hat{a}_\al(in) \hat{S}'^\La_{t_{out},t_{in}}.
\end{gathered}
\end{equation}
The annihilation operators $\hat{a}_\al(in)$ and $\hat{a}_\al(out)$ must act in the same Fock space $F_{t_{in}}$, and their vacuum states are
\begin{equation}
    |\overline{in}\ran:=|vac,t_{in}\ran\in F_{t_{in}},\qquad |\overline{out}\ran:=\hat{U}^\La_{t_{in},t_{out}}|vac,t_{out}\ran\in F_{t_{in}}.
\end{equation}
The matrix element of the evolution operator is given by
\begin{equation}
    \lan vac,t_{out}|\hat{U}^\La_{t_{out},t_{in}}|vac,t_{in}\ran=\lan \overline{out}|\overline{in}\ran.
\end{equation}
Let
\begin{equation}\label{Heis_eqs}
    \hat{Z}^A(t):=\hat{U}^\La_{t_{in},t}\hat{Z}^A\hat{U}^\La_{t,t_{in}},\qquad i\dot{\hat{Z}}^A(t)=[\hat{Z}^A(t),\hat{H}'_\La(t)],
\end{equation}
where $\hat{H}'_\La(t)$ is \eqref{Hamilt_reg1} written in the Heisenberg representation. From \eqref{ZinAAd}, we obtain
\begin{equation}\label{Zoutin}
\begin{split}
    \hat{Z}^A(t_{out})&=-i\sum_\al\big[\ups_\al^A(t_{out})\hat{a}_\al(out)-\bar{\ups}_\al^A(t_{out})\hat{a}^\dag_\al(out)\big],\\
    \hat{Z}^A(t_{in})&=-i\sum_\al\big[\ups_\al^A(t_{in})\hat{a}_\al(in)-\bar{\ups}_\al^A(t_{in})\hat{a}^\dag_\al(in)\big].
\end{split}
\end{equation}
On the other hand, introducing the commutator Green function
\begin{equation}
    \tilde{G}^{AB}_\La(t,t'):=[\hat{Z}^A(t),\hat{Z}^B(t')],
\end{equation}
and using the commutation relations \eqref{Hamilt_quadr}, we can write
\begin{equation}\label{Zout}
    \hat{Z}^A(t_{out})=-i\tilde{G}^A_{\La B}(t_{out},t_{in}) \hat{Z}^B(t_{in}),\qquad \tilde{G}^A_{\La B}(t_{out},t_{in}):= \tilde{G}^{AC}_\La(t_{out},t_{in}) J_{CB}.
\end{equation}
Notice that $\tilde{G}^{AB}_{\La}(t,t')$ is a $c$-number. It is clear that $\tilde{G}^{AB}_{\La}(t,t')$ tends to the commutator Green function associated with the initial classical Hamiltonian $H(t)$ in the regularization removal limit. It follows from the relations \eqref{creaannh_oper}, \eqref{Zoutin}, \eqref{Zout} that
\begin{equation}\label{canon_trans_expl}
\begin{split}
    \hat{a}_\al(out)&=-\bar{\ups}_\al^A(t_{out})\tilde{G}^\La_{AB}(t_{out},t_{in})\ups^B_\be(t_{in})\hat{a}_\be(in) +\bar{\ups}_\al^A(t_{out})\tilde{G}^\La_{AB}(t_{out},t_{in})\bar{\ups}^B_\be(t_{in})\hat{a}^\dag_\be(in),\\
    \hat{a}^\dag_\al(out)&=-\ups_\al^A(t_{out})\tilde{G}^\La_{AB}(t_{out},t_{in})\ups^B_\be(t_{in})\hat{a}_\be(in) +\ups_\al^A(t_{out})\tilde{G}^\La_{AB}(t_{out},t_{in})\bar{\ups}^B_\be(t_{in})\hat{a}_\be(in),
\end{split}
\end{equation}
where $\tilde{G}^\La_{AB}=J_{AC}\tilde{G}^C_{\La B}$ and summation over all $\be$ is understood. Let us stress that the dependence of the mode functions on $t_{in}$, $t_{out}$ in this expression does not stem from the evolution but is determined by the solution of the problem \eqref{eigen_prblm}. The mode functions depend on $t$ only through the background fields taken at the instant of time $t$.

The linear canonical transform \eqref{canon_trans_expl} is a unitary one if and only if
\begin{equation}\label{Psi_def}
    \Psi_{\al\be}:=\bar{\ups}_\al^A(t_{out})\tilde{G}^\La_{AB}(t_{out},t_{in})\bar{\ups}^B_\be(t_{in})
\end{equation}
is Hilbert-Schmidt (do not confuse $\Psi_{\al\be}$ with $i\bar{G}_{\al\be}$ defined in \eqref{FandG}). As follows from \eqref{CAO_Heis}, this is a necessary and sufficient condition for unitarity of $\hat{S}'^\La_{t_{out},t_{in}}$. It was shown in \cite{Scharf79} for the Dirac fermions evolving on good electromagnetic backgrounds that the respective operator $\Psi_{\al\be}$ is Hilbert-Schmidt in the regularization removal limit.

It is well known that
\begin{equation}\label{part_creat}
    n_\al=\sum_\be|\Psi_{\al\be}|^2
\end{equation}
is the average number of particles in the $out$-state $\al$ created from the $in$-vacuum state during the evolution. In the next section, we shall find the leading ultraviolet asymptotics of this number for a scalar field on a time-dependent metric background in the regularization removal limit.

\section{QFT with respect to a congruence of observers}\label{Scal_Field}

Let us consider a congruence of observers in the spacetime $M$ with the metric
\begin{equation}
    \eta_{ab}=diag(1,-1,-1,-1).
\end{equation}
The congruence of observers is standardly described (see, e.g., \cite{DeWQG1,Vladimir.12,Mitskev.12,KucharI,KucharII,Isham.12,KucharPoT,Taub,FockB}) by an orientation preserving smooth map $\vk^a(x)$, $a=\overline{0,3}$, $i=\overline{1,3}$, of the four-dimensional manifold $N$ to $M$. The observers manifold $N$ is equipped with the vector field $\xi^\mu(x)$, $\mu=\overline{0,3}$, such that $\xi^\mu g_{\mu\nu}\xi^\nu>0$, where $g_{\mu\nu}:=\partial_\mu \vk^a\partial_\nu \vk^b \eta_{ab}$ is the induced metric on $N$. If we set $\xi^\mu=(1,0,0,0)$, these geometric constructions have a clear physical interpretation. The map $\vk^a(x)$ with fixed $x^i$ describes the worldline of a given observer, $x^i$ parameterize observers in the congruence, and $x^0\equiv t$ is the time of a given observer. The induced metric $g_{\mu\nu}(x)$ is the metric that is measured by the congruence of observers with the aid of radiolocation (a thorough description of this method can be found, for example, in \cite{LandLifshCTF}). In this sense, $g_{\mu\nu}$ can be called as the result of a classical measurement of the spacetime metric by the congruence of observers \cite{DeWQG1}. This measurement is classical since we do not take into account the influence of observers on the metric and the quantum recoil experienced by them.

In order to allow for these quantum effects, we have to introduce the quantum fields into the theory that describe the observers (if these fields are not already present), quantize somehow the metric field, and consider the states that take into account the presence of the observers, which, of course, are made of particles. However, there is an issue in realization of such a program. There is no natural candidate $\xi^\mu[\Phi]$ or its quantum counterpart $\hat{\xi}^\mu[\Phi]$ for the role of the timelike vector field characterizing the congruence of observers. Here, $\xi^\mu[\Phi]$ should be some local in time functional of the fields of the standard model with gravity. In quantum gravity this problem is known as the problem of time (see for review, e.g., \cite{Isham.12,KucharPoT}). It may well happen that the observables of quantum theory would be independent of the choice of $\xi^\mu[\Phi]$, i.e., it does not matter whether one uses a clock on a wall or a laptop clock, which tick non-uniformly with respect to each other, to describe the results of the experiments. However, the results we shall obtain below suggest that this is not the case at a one-loop level, at least. Quantum field theory depends on the choice of $\xi^\mu[\Phi]$ \cite{gmse.11,KalKaz1,KalKaz2.12,qgadm,KazMil1} and becomes non-unitary in the regularization removal limit for some ``unfortunate'' choices of this vector field when the spacetime dimension $D\geqslant3$. In fact, we shall show that at a one-loop level, where the quantum fields freely propagate on a given classical background, the quantum evolution is not unitary in the regularization removal limit for a general choice of the congruence of observers. A similar conclusion was drawn in \cite{TorVara} for the different splitting of a quantum field into positive- and negative-frequency parts. The results of \cite{Park1,FullingAQFT,GriMaMos.11,ParkTom} for cosmological metrics can also be interpreted in this way. It is noteworthy to stress once again that the results we shall obtain does not mean that the quantum evolution is not unitary for any Hamiltonian and any time dependent metric background when $D\geqslant3$. For example, it was shown in \cite{Pavl,GriPav} that for the FLRW metric in the conformal coordinates one can construct such a Hamiltonian for a rescaled scalar field that the number of created particles defined by its diagonalization is finite. The asymptotics we shall find holds for metrics of a general form written with respect to a general congruence of observers. It is these metrics and congruences of observers that are realized in Nature.

To simplify the problem, we consider the evolution of a quantum massive scalar field in the Minkowski spacetime $M$ with respect to the congruence of observers. The perturbation of the metric caused by the observers is assumed to be negligible, and we shall see that these perturbations do not alter the main conclusion. We also disregard the dynamics of other quantum fields since they evolve independently at a one-loop level. The violation of unitarity for one quantum field ruins unitarity of a whole system. As for the massive scalar field, we shall find the explicit formulas for the ultraviolet asymptotics of average number of created particles \eqref{part_creat} and show that, for a general congruence of observers, the sum over $\al$ diverges for $D\geqslant3$, i.e., $\Psi_{\al\be}$ is not Hilbert-Schmidt in the regularization removal limit.

The action for a massive scalar field on the background with the metric $g_{\mu\nu}$ has the standard form
\begin{equation}
    S[\phi]=\frac12\int d^Dx\sqrt{|g|}(\partial_\mu\phi g^{\mu\nu}\partial_\nu\phi-m^2\phi^2).
\end{equation}
The metric components measured by the congruence of observers satisfy the constraints \cite{LandLifshCTF}:
\begin{equation}\label{metric_constr}
    g_{00}>0,
\end{equation}
and $g^{ij}$ is negative-definite. Further, we shall need the following relations \cite{LandLifshCTF}
\begin{equation}\label{g00_neg}
    \bar{g}_{ik}g^{kj}:=\de_i^j,\qquad g_{00}\det\bar{g}_{ij}=\det g_{\mu\nu},\qquad g^{00}- g^{ij} g_ig_j=(g_{00})^{-1},\qquad g_i=-\bar{g}_{ij}g^{j0},
\end{equation}
where $g_i:=g_{0i}/g_{00}$. Let us introduce the canonical momentum
\begin{equation}\label{Legandre_trans}
    \pi:=\frac{\partial \mathcal{L}}{\partial\dot{\phi}}=\sqrt{|g|}(g^{00}\dot{\phi}+g^{0i}\partial_i\phi),
\end{equation}
where $\dot{\phi}=\partial_t\phi$ and $\mathcal{L}$ is the Lagrangian density. Then the Hamiltonian density
\begin{equation}\label{Hamilt_scal}
\begin{split}
    \mathcal{H}=\pi\dot{\phi}-\mathcal{L}=&\frac12\Big[\frac{\pi^2}{g^{00}\sqrt{|g|}} -2\frac{\pi g^{0i}\partial_i\phi}{g^{00}} -\sqrt{|g|}(\tilde{g}^{ij}\partial_i\phi\partial_j\phi-m^2\phi^2) \Big]\\
    =&\frac12\Big[\frac{(\pi-\sqrt{|g|}g^{0i}\partial_i\phi)^2}{g^{00}\sqrt{|g|}}-\sqrt{|g|}(g^{ij}\partial_i\phi\partial_j\phi-m^2\phi^2 )\Big],
\end{split}
\end{equation}
where $\tilde{g}^{ij}=g^{ij}-g^{0i}g^{0j}/g^{00}=(g_{ij})^{-1}$. If $g^{00}>0$ and $g^{ij}$ is negative-definite, then the corresponding quadratic form $H_{AB}$ is positive-definite. These conditions are necessary and sufficient for positive definiteness of $H_{AB}$. Notice that the conditions \eqref{metric_constr} do not imply that $g^{00}>0$, i.e., the hypersurfaces $t=const$ can be non-spacelike. The inequality \eqref{metric_constr} is satisfied when there is a transversal future-directed timelike vector at every point of the hypersurface $t=const$. In terms of the normal covector $N_\mu=(1,0,0,0)$ to the hypersurface $t=const$, the conditions \eqref{metric_constr} say that it must not lie inside of the past light cone. Therefore, the quadratic form $H_{AB}$ can be indefinite for a classically proper congruence of observers, and we do not exclude, in advance, such a possibility.

The quadratic form $H_{AB}$ is written as
\begin{equation}
    H_{AB}=\left[
             \begin{array}{cc}
               \partial_i\sqrt{|g|}\tilde{g}^{ij}\partial_j+\sqrt{|g|}m^2 & \partial_i\frac{g^{i0}}{g^{00}} \\
               -\frac{g^{0i}}{g^{00}}\partial_i & \frac{1}{g^{00}\sqrt{|g|}} \\
             \end{array}
           \right].
\end{equation}
The eigenvalue problem \eqref{eigen_prblm} takes the form
\begin{equation}\label{eigen_prblm_scal}
    \left[
       \begin{array}{c}
         \partial_i(\sqrt{|g|}\tilde{g}^{ij}\partial_ju_\al)+\sqrt{|g|}m^2u_\al+\partial_i(\frac{g^{i0}}{g^{00}}w_\al) \\
         -\frac{g^{0i}}{g^{00}}\partial_iu_\al+\frac{w_\al}{g^{00}\sqrt{|g|}} \\
       \end{array}
     \right]=i\omega_\al\left[
                          \begin{array}{c}
                            w_\al \\
                            -u_\al \\
                          \end{array}
                        \right],\qquad\ups^A_\al(t)=\left[
                                                   \begin{array}{c}
                                                     u_\al(t) \\
                                                     w_\al(t) \\
                                                   \end{array}
                                                 \right].
\end{equation}
Combining these expressions, we come to the equations
\begin{equation}\label{KG_stat}
\begin{split}
    \Big[(\hat{p}_i+\omega_\al g_i)\sqrt{|g|}g^{ij}(\hat{p}_j+\omega_\al g_j)+\sqrt{|g|}\big(\frac{\omega^2_\al}{g_{00}}-m^2\big)\Big]&u_\al=0,\\
    w_\al=-i\sqrt{|g|}\Big[g_i g^{ij}(\hat{p}_j+\omega_\al g_j)+\frac{\omega_\al}{g_{00}}\Big]&u_\al,
\end{split}
\end{equation}
where $\hat{p}_i=-i\partial_i$. Let $H(\omega)$ be the operator standing on the left-hand side of the first equation in \eqref{KG_stat}. We assume that the metric components are infinitely smooth, the constraints \eqref{metric_constr} are fulfilled, and the problem \eqref{KG_stat} is posed in a large box with the side $L$ with the Dirichlet boundary conditions. Then $H(\omega)$ is self-adjoint with respect to the standard scalar product
\begin{equation}
    \lan u_1|u_2\ran=\int d\spx\bar{u}_1(\spx)u_2(\spx).
\end{equation}
Denoting by $\e_\al(\omega)$ the eigenvalues of $H(\omega)$ corresponding to the eigenvectors $u_\al(\omega)$, we have
\begin{equation}
\begin{split}
    \e'_\al(\omega)\lan u_\al(\omega)|u_\al(\omega)\ran&=\lan u_\al(\omega)|H'(\omega)u_\al(\omega)\ran\\
    &=\int d\spx\sqrt{|g|}\Big[\bar{u}_\al\big(g^i \hat{P}_i(\omega)+\frac{\omega}{g_{00}}\big)u_\al +u_\al\overline{\big(g^i \hat{P}_i(\omega)+\frac{\omega}{g_{00}}\big)u_\al} \Big],
\end{split}
\end{equation}
where $g^i:=g^{ij}g_j$, $\hat{P}_i(\omega):=\hat{p}_i+\omega g_i$. Taking $\omega=\omega_\al$, where $\e_\al(\omega_\al)=0$, we obtain
\begin{equation}\label{eprime}
    \e'_\al(\omega_\al)\lan u_\al(\omega_\al)|u_\al(\omega_\al)\ran=i\{\ups_\al,\bar{\ups}_\al\}=\omega_\al^{-1}\bar{\ups}^A_\al H_{AB}\ups^B_\al.
\end{equation}
In accordance with \eqref{eigen_prblm}, we choose the splitting of the mode functions into positive-frequency and negative-frequency ones such that $\e'_\al(\omega_\al)>0$ for the positive-frequency modes \cite{MigdalMM,Furshep,FursVass,KalKaz1,KalKaz2.12,KalKaz3}.

Now we show that if there are regions where $g^{00}<0$, then, in general, the mode functions do not constitute a complete set \eqref{orth_complt}. In virtue of the third relation in \eqref{g00_neg} and inequality \eqref{metric_constr}, $g^{00}>0$ when $g_i$ is small. Let us choose
\begin{equation}
    g_i=\tilde{g}_i+\lambda\partial_i\vf,
\end{equation}
where $\lambda\in[0,1]$, the function $\vf(x)$ is infinitely smooth with a compact support, and
\begin{equation}
    g_{00}^{-1}+g^{ij}\tilde{g}_i\tilde{g}_j>0,
\end{equation}
while for $\la=1$ there are domains where $g^{00}<0$. We also suppose $\tilde{g}_i$ to be sufficiently general such that the eigenvalues $\e_\al(\omega,\la)$ of $H(\omega,\la)$ are nondegenerate at $\la=0$. This implies that the modes corresponding to different $\al$ are orthogonal as in \eqref{orth_complt}. The substitution
\begin{equation}\label{u_scal}
    u_\al=e^{-i\la\omega_\al\vf}f_\al
\end{equation}
into the first equation in \eqref{KG_stat} shows that the spectra $\e_\al(\omega)$, $\omega_\al$ and the functions $f_\al$ do not depend on $\la$, the scalar product
\begin{equation}
    \lan u_\al|u_\al\ran=\lan f_\al|f_\al\ran
\end{equation}
is independent of $\la$ too, and all the eigenvectors \eqref{eigen_prblm_scal} are of the form \eqref{u_scal} with $w_\al$ given by the second equation in \eqref{KG_stat}. Then it follows from \eqref{eprime} that
\begin{equation}
    \omega_\al^{-1}\bar{\ups}^A_\al H_{AB}\ups^B_\al
\end{equation}
does not depend on $\la$ and positive. However, $\omega_\al$ does not change with $\la$ and so
\begin{equation}
    \omega_\al>0,\qquad \bar{\ups}^A_\al H_{AB}\ups^B_\al>0,
\end{equation}
for $\la\in[0,1]$. As the orthogonality property of the modes corresponding to different $\al$ holds for the indefinite $H_{AB}$ as well, they remains orthogonal for $\la=1$. Therefore, the first two relations in \eqref{orth_complt} are fulfilled for $\la=1$. Suppose that the completeness relation \eqref{orth_complt} also holds. Then $Z^A$ can be written in the form \eqref{ZinAAd} with the creation-annihilation operators replaced by the complex functions $a_\al(t)$ and $\bar{a}_\al(t)$. Substituting that representation to the Hamiltonian, we come to \eqref{Hamilton_diagonalized}, which is a positive-definite expression. On the other hand, taking $\phi(x)=0$ and $\pi(x)$ to be infinitely smooth with a compact support in the region where $g^{00}<0$, we find that
\begin{equation}
    \frac12Z^A H_{AB}Z^B<0.
\end{equation}
This contradiction shows that the mode functions cannot be complete in this case.

Of course, the violation of the completeness relation gives rise to the violation of unitarity of the quantum evolution described in Sec. \ref{Hamilt_Diag} when one starts from the spacelike hypersurface $t=t_{in}$ and finishes the evolution on the non-spacelike hypersurface $t=t_{out}$. This shows that quantum theory does depend on the choice of the congruence of observers and becomes non-unitary for non-spacelike hypersurfaces of an equal time even at a finite energy cutoff. We draw to a conclusion that not all the classical proper congruences of observers are admissible in a unitary quantum theory. Below, we suppose that all the equal time hypersurfaces are spacelike and so $H_{AB}$ is positive-definite and $\omega_\al>0$.

Since our goal is to find the ultraviolet asymptotics of \eqref{part_creat} in the regularization removal limit $\La\rightarrow+\infty$, we may substitute the regularized commutator Green function $\tilde{G}^\La(t_{out},t_{in})$ by its regularization removal limit. In this limit (see, e.g., \cite{KazMil1}),
\begin{equation}\label{comm_Gr_fun}
    \tilde{G}_{AB}(x^0,y^0)=
    \left[
      \begin{array}{cc}
        -\sqrt{|g(x)|}g^{0\mu}(x)\sqrt{|g(y)|}g^{0\nu}(y)\partial^{xy}_{\mu\nu} & \sqrt{|g(x)|}g^{0\mu}(x)\partial_\mu^x \\
        \sqrt{|g(y)|}g^{0\mu}(y)\partial_\mu^y & -1 \\
      \end{array}
    \right]i\tilde{G}(x,y),
\end{equation}
where $\tilde{G}(x,y)$ is the commutator Green function for a massive scalar field in the Minkowski spacetime written in the curvilinear coordinates (see, e.g., \cite{BogolShir})
\begin{equation}
    \tilde{G}(x,y)=\int\frac{d\spk}{(2\pi)^d}\Big[\frac{e^{ik_a(\vk^a(x)-\vk^a(y))}}{2ik_{\ddo}} -\frac{e^{-ik_a(\vk^a(x)-\vk^a(y))}}{2ik_{\ddo}}\Big],\qquad k_{\ddo}:=\sqrt{\spk^2+m^2},
\end{equation}
where $d:=D-1$ and we introduce the notation $\ddo$ for the index $a=0$ to distinguish it from $\mu=0$. Note, in passing, that there is no any problem with the classical evolution generated by \eqref{comm_Gr_fun} from one hypersurface satisfying \eqref{metric_constr} to another one even in the case when there exist domains on these hypersurfaces where $g^{00}<0$, provided this evolution starts with the smooth initial data $\phi(t_{in},\spy)$, $\dot{\phi}(t_{in},\spy)$. Substituting \eqref{comm_Gr_fun} into \eqref{Psi_def}, we obtain
\begin{equation}\label{Psi_bar}
\begin{split}
    \bar{\Psi}_{\al\be}&=-\ups_\al^A(x^0)\tilde{G}_{AB}(x^0,y^0)\ups^B_\be(y^0)=\\
    &=\int\frac{d\spx d\spk d\spy}{(2\pi)^d}\sqrt{gg'} \bigg\{\Big[\big(k^0+g^i \hat{P}_i(\omega_\al)+\frac{\omega_\al}{g_{00}}\big) u_\al\Big] \Big[\big(k'^0-g'^i \hat{P}'_i(\omega_\be)-\frac{\omega_\be}{g'_{00}}\big) u'_\be\Big]\\
    &\times\frac{e^{ik_a(\vk^a-\vk'^a)}}{2k_{\ddo}}+(k_a\leftrightarrow-k_a)\bigg\},
\end{split}
\end{equation}
where the quantities without primes correspond to the point $x=(x^0,\spx)$ and the quantities with primes refer to the point $y=(y^0,\spy)$. Also we denote as
\begin{equation}
    k^0=k^0(x)=g^{0\mu}(x)e^a_\mu(x)k_a,\qquad e^a_\mu(x):=\partial_\mu\vk^a(x),
\end{equation}
and the analogous expression for $k'^0$. In the next section, we shall derive the ultraviolet asymptotics of \eqref{Psi_bar} under the assumption that
\begin{equation}\label{comp_supp}
    g_{\mu\nu}(x)=\eta_{\mu\nu},\qquad e^a_\mu(x)=\de^a_\mu
\end{equation}
for all $x^0$ and $|\spx|>R$, where $R$ is much smaller than the side $L$ of a large box where the problem \eqref{KG_stat} is posed.

\section{Ultraviolet asymptotics}\label{ULtr_Asympt}

In order to find the ultraviolet asymptotics of \eqref{Psi_bar}, we need to obtain the expression for the mode functions $u_\al(\omega_\al)$ at large $\omega_\al$. This corresponds to the short-wave (or WKB) approximation to the solutions of the first equation in \eqref{KG_stat}. The procedure of how to find the expansion of $u_\al(\omega)$ in the asymptotic series in $\omega_\al^{-1}$ is well known (see, e.g., \cite{BabBuld,MaslovCWKB,BBT.1,BBTZ}). Notice that the first equation in \eqref{KG_stat} describes the stationary solutions to the Klein-Gordon equation with the stationary metric obtained from $g_{\mu\nu}(x)$ by freezing the variable $x^0$. This stationary metric is not flat, in general, despite the fact that $\eta_{ab}$ is flat.

In accordance with the general procedure, we seek for the solution of the form
\begin{equation}\label{WKB_form}
    u_\al=h_\al e^{iS_\al}.
\end{equation}
Then, equation \eqref{KG_stat} becomes
\begin{multline}
    \Big[\sqrt{|g|}\Big(g^{ij}P^\al_iP^\al_j+\frac{\omega^2_\al}{g_{00}}\Big)+P^\al_i\sqrt{|g|}g^{ij}\hat{p}_j+\hat{p}_i\sqrt{|g|} g^{ij}P^\al_j \\
    +\hat{p}_i\sqrt{|g|} g^{ij}\hat{p}_j-\sqrt{|g|}m^2 \Big](h_\al^{(0)}+h_\al^{(1)}+\cdots)=0,
\end{multline}
where $P^\al_i:=\partial_iS_\al+\omega_\al g_i=p^\al_i+\omega_\al g_i$, and all the terms are arranged according to their power $\omega_\al^{-1}$. The leading order term is nullified by solving the Hamilton-Jacobi equation
\begin{equation}\label{HJ_eqs}
    g^{ij}P^\al_iP^\al_j+\frac{\omega^2_\al}{g_{00}}=0.
\end{equation}
For $g^{00}>0$, the generalized $P_i^\al$ and kinetic $p_i^\al$ momenta satisfying the Hamilton-Jacobi equation are in one-to-one correspondence, because
\begin{equation}
    \omega_\al=(g^{00})^{-1}\big[-g^ip_i^\al+\sqrt{(g^ip_i^\al)^2-g^{00}g^{ij}p_i^\al p_j^\al}\big]>0.
\end{equation}
The Hamiltonian for \eqref{HJ_eqs} is
\begin{equation}\label{Hamiltonian}
    H=\frac12\Big[g^{ij}(p_i+\omega_\al g_i)(p_j+\omega_\al g_j)+\frac{\omega^2_\al}{g_{00}}\Big],
\end{equation}
and the respective Hamilton equations read as
\begin{equation}\label{Ham_eqs}
\begin{split}
    \dot{x}^i&=g^{ij}(p_j+\omega_\al g_j),\\
    \dot{p}_i&=\frac12\partial_i g_{00}\frac{\omega_\al^2}{g^2_{00}} -\omega_\al g^{kl}\partial_i g_k (p_l+\omega_\al g_l) -\frac12\partial_ig^{kl} (p_k+\omega_\al g_k)(p_l+\omega_\al g_l).
\end{split}
\end{equation}
These are the equations of motion of a massless particle on a stationary metric background.

Consider a uniform flux of particles  starting from the plane, which lies outside of the ball $|\spx|\leqslant R$, with momenta $\spp_\al$ orthogonal to this plane. In the ray coordinates $(\tau,\s)$, where $\tau$ is the length counted along the ray with the help of the metric $-\bar{g}_{ij}$ and $\s$ are the transversal coordinates (see, for details, \cite{BabBuld}), we have
\begin{equation}
    S_\al(\tau,\s)=-\omega_\al\int_0^\tau d\tau'\big[g^{-1/2}_{00}(\tau',\s)+g_\tau(\tau',\s)\big],\qquad \omega_\al=|\spp_\al|,
\end{equation}
and
\begin{equation}\label{h0}
    h^{(0)}_\al=\bigg[\frac{g^{1/2}_{00}(\tau,\s)\vf(\s)}{2\omega_\al\sqrt{|g(\tau,\s)|}}\bigg]^{1/2},
\end{equation}
where $\vf(\s)$ is an arbitrary function. In the initial coordinates,
\begin{equation}\label{WKB_Jack}
    \sqrt{|g(\tau,\s)|}=\sqrt{|g(x)|}\det(\partial x^i/\partial(\tau,\s)).
\end{equation}
This expression vanishes on caustics where the above procedure does not work. It can be improved by the standard means (see, e.g., \cite{MaslovCWKB,BBTZ}). The caustics are supported on the set of measure zero and their contribution to the integral \eqref{Psi_bar} can be neglected since the singularity of \eqref{h0} is integrable there. The next orders $h^{(k)}_\al$ of the expansion can readily be found in the ray coordinates. However, their contribution to \eqref{Psi_bar} is subleading and we disregard them.

Let us check that the mode functions
\begin{equation}
    \ups_\al^{(0)}=\left[
                     \begin{array}{c}
                       1 \\
                       -i\sqrt{|g|}(g^i\hat{P}_i(\omega_\al)+\omega_\al/g_{00}) \\
                     \end{array}
                   \right]u_\al^{(0)},\qquad u_\al^{(0)}=h_\al^{(0)}e^{iS_\al},
\end{equation}
are orthogonal in the leading order in $\omega_\al^{-1}$ and normalize them. The modes corresponding to different energies $\omega_\al$ are orthogonal in virtue of the general properties or the mode functions. In fact, we only need to consider the skew-symmetric product between $\ups^{(0)}_\al$ and $\bar{\ups}^{(0)}_\be$ for the same energy $\omega_\al=\omega_\be$. In this case,
\begin{equation}\label{normalization}
    \{\ups_\al,\bar{\ups}_\be\}\approx-i\int d\spx\sqrt{|g|}\Big[g^i(P^\al_i+P^\be_i)+\frac{2\omega_\al}{g_{00}}\Big] u_\al^{(0)}\bar{u}_\be^{(0)}.
\end{equation}
This integral can be evaluated for large $\omega_\al$ by means of the WKB method. The stationary points are found from the equation
\begin{equation}
    p^\al_i=p^\be_i.
\end{equation}
This implies that ($\omega_\al=\omega_\be$)
\begin{equation}
    P^\al_i=P^\be_i
\end{equation}
at the same point $\spx$. Therefore, moving along the ray to the past, we conclude that either $\spp_\al=\spp_\be$ or there is no stationary point. In the latter case, the integral \eqref{normalization} vanishes in the leading order and we are left with the case $\spp_\al=\spp_\be$. In this case, it is convenient to evaluate the integral \eqref{normalization} in the ray coordinates. Bearing in mind that
\begin{equation}
    \frac{\omega_\al}{g_{00}} +g^iP^\al_i=\omega_\al g_{00}^{-1/2}(g_{00}^{-1/2}+g_\tau),
\end{equation}
we obtain
\begin{equation}\label{normal_app}
    \{\ups_\al,\bar{\ups}_\al\}\approx-i\int d\tau d\s\vf(\s)(g_{00}^{-1/2}+g_\tau)= \frac{i}{\omega_\al}\int d\s\vf(\s)S_\al(\infty,\s)=-i,
\end{equation}
where $S_\al(\infty,\s)$ is the value of the Hamilton-Jacobi action along the ray. Taking
\begin{equation}
    \vf(\s)=-\frac{\omega_\al}{L^2 S_\al(\infty,\s)},
\end{equation}
we completely specify the approximate mode functions $u_\al^{(0)}$.

The normalization factor can be simplified in view of the conditions imposed in \eqref{comp_supp}. The integrand of \eqref{normal_app} equals
\begin{equation}
    -\vf(\s)\omega_\al L
\end{equation}
for $\s$ lying outside of the cross-section of the ball $|\spx|\leqslant R$. As long as the side $L$ of the box, where the problem is posed, is much larger than $R$, the integral \eqref{normal_app} can be written as
\begin{equation}
    \{\ups_\al,\bar{\ups}_\al\}\approx -i L\int d\s\vf(\s) (1+O(R^2/L^2))=-i,
\end{equation}
and so
\begin{equation}
    \vf(\s)\approx\frac{1}{L^3}\equiv\frac{1}{V}.
\end{equation}
Thus we have
\begin{equation}\label{WKB_norm}
    h_\al^{(0)}\approx\bigg[\frac{g^{1/2}_{00}(\tau,\s)}{2\omega_\al V\sqrt{|g(\tau,\s)|}}\bigg]^{1/2}=\big[2\omega_\al V\sqrt{|\bar{g}(\tau,\s)|}\big]^{-1/2},
\end{equation}
where $\bar{g}:=\det\bar{g}_{ij}$. Notice that the mass $m$ of the particle does not enter into the expression for $\ups_\al^{(0)}$. These approximate mode functions describe massless particles as it should be in the ultraviolet limit.

On substituting the representation \eqref{WKB_form} into \eqref{Psi_bar}, we have
\begin{equation}\label{Psi_bar_WKB}
\begin{split}
    \bar{\Psi}_{\al\be} =&\int\frac{d\spx d\spk d\spy}{(2\pi)^d}\sqrt{gg'} \bigg\{\Big[\big(k^0+g^i P^\al_i+\frac{\omega_\al}{g_{00}} +g^i\hat{p}_i\big) h_\al\Big]
    \Big[\big(k'^0-g'^i P'^\be_i -\frac{\omega_\be}{g'_{00}} -g'^i\hat{p}'_i\big) h'_\be\Big]\\ &\times\frac{e^{ik_a(\vk^a-\vk'^a)+iS_\al+iS'_\be}}{2k_{\ddo}}
    +(k_a\leftrightarrow-k_a)\bigg\},
\end{split}
\end{equation}
We shall expand this integral into an asymptotic series with respect to the small parameter $\omega^{-1}\sim\omega^{-1}_\al\sim\omega^{-1}_\be$ by the WKB method. To trace the orders in $\omega^{-1}$ correctly, one has to stretch the integration variable $\spk\rightarrow \omega \spk$. However, we shall just keep in mind that $\spk$ is of the order  $\omega$. Then we single out the leading in $\omega$ terms in the exponent entering into the integrand of \eqref{Psi_bar_WKB} and find the stationary points of the resulting expression. As for the first term in the curly brackets in \eqref{Psi_bar_WKB}, we obtain
\begin{equation}\label{stat_p1}
    p'^\be_i=q_a e'^a_i,\qquad p_i^\al=-q_a e^a_i,\qquad \vk^{\bar{a}}-\vk'^{\bar{a}}=-n_{\bar{a}}(\vk^{\ddo}-\vk'^{\ddo}),
\end{equation}
where $n_{\bar{a}}=q_{\bar{a}}/q_{\ddo}$, $q_a=(|\spk|,k_{\bar{a}})$, and $\bar{a}=\overline{1,3}$. The corrections due to mass  entering into $k_{\ddo}$ are of the order $\omega^{-2}$ in comparison with the leading contribution. As we shall see, these can safely be neglected.

The first two equations in \eqref{Psi_bar_WKB} can be resolved with respect to $q_a$ as
\begin{equation}\label{light_v}
\begin{aligned}
    q_a&=e'^\mu_a\pi'_\mu,&\qquad \pi'_\mu&=\big(\omega_\be+\frac{2}{g'^{00}}g'^ip'^\be_i,p'^\be_i\big),\\
    q_a&=-e^\mu_a\pi_\mu,&\qquad \pi_\mu&=\big(-\omega_\al,p^\al_i\big).
\end{aligned}
\end{equation}
These solutions are unique when $g^{00}>0$. Substituting one of these solutions to the equations in \eqref{stat_p1}, we obtain a one-to-one correspondence between $p^\al_i$ and $p'^\be_i$. The third equation in \eqref{stat_p1} says that the point $x$ belongs to a light ray emanated in the spacetime from the point $y$ along the unit vector $n^{\bar{a}}$. This provides the following geometrical picture of how to construct the solution to \eqref{stat_p1}. Take some point $y$ and future directed light-like covector $q_a$; emanate the light ray from the point $y$ along $n^{\bar{a}}$ to the point $x$ belonging to the hypersurface with given $x^0$; take $p_i^\al$ at the point $x$ and $p'^\be_i$ at the point $y$ as given by \eqref{stat_p1} and, moving to the past along the solutions of the Hamilton equations \eqref{Ham_eqs}, obtain $\spp_\al$ and $\spp_\be$, respectively. This construction gives the map of a $2d$ dimensional manifold with the coordinates $(\spy,q_{\bar{a}})$ to $2d$ dimensional linear space of $(\spp_\be,\spp_\al)$, which is nondegenerate, in a general position. It means that, in a general position, there is, at most, a discrete set of points $(\spy,q_{\bar{a}})$ (or, equivalently, $(\spy,\spx)$) that satisfy \eqref{stat_p1} for fixed $(\spp_\be,\spp_\al)$. Furthermore, equations \eqref{HJ_eqs}, \eqref{stat_p1} are homogeneous with respect to momenta and so if a solution to \eqref{stat_p1} exists for some $(\spp_\be,\spp_\al)$ then there exists a solution for $(\la\spp_\be,\la\spp_\al)$, $\la>0$. In other words, the set of $(\spp_\be,\spp_\al)$ admitting a solution to \eqref{stat_p1} is a conical neighborhood.

Now we turn to the second term in the curly brackets in \eqref{Psi_bar_WKB}. The stationary points are determined by the equations
\begin{equation}\label{stat_p2}
    p'^\be_i=-q_a e'^a_i,\qquad p_i^\al=q_a e^a_i,\qquad \vk^{\bar{a}}-\vk'^{\bar{a}}=-n_{\bar{a}}(\vk^{\ddo}-\vk'^{\ddo}),
\end{equation}
which are resolved with respect to $q_a$ as
\begin{equation}
\begin{aligned}
    q_a&=-e'^\mu_a\pi'_\mu,&\qquad \pi'_\mu&=\big(-\omega_\be,p^\be_i\big),\\
    q_a&=e^\mu_a\pi_\mu,&\qquad \pi_\mu&=\big(\omega_\al+\frac{2}{g^{00}}g^ip^\al_i,p^\al_i\big).
\end{aligned}
\end{equation}
Applying the above considerations to construct a solution to \eqref{stat_p2}, one can see that, in a general position, the stationarity conditions \eqref{stat_p1} and \eqref{stat_p2} are fulfilled for different points $x$ and $y$ at the fixed $\spp_\al$ and $\spp_\be$. Therefore, the two terms in the curly brackets in \eqref{Psi_bar_WKB} cannot cancel each other.

The stationarity condition \eqref{stat_p1} implies that
\begin{equation}
    q_a(\vk^a-\vk'^a)=0,\qquad k^0=g^i P^\al_i+\frac{\omega_\al}{g_{00}}=g'^i P'^\be_i+\frac{\omega_\be}{g'_{00}},
\end{equation}
and the same relations hold for the stationary points \eqref{stat_p2}. Therefore, the naively expected leading order contribution to $\bar{\Psi}_{\al\be}$ vanishes. One has to expand the preexponential factor and the expression standing in the exponent in \eqref{Psi_bar_WKB} in a Taylor series near the stationary points and evaluate the resulting Gaussian integrals (the standard WKB procedure). The following estimates take place
\begin{equation}
    \lan\De x^i\De x^j\ran\sim\omega^{-1-d/2},\qquad \lan\De x^i\De k_{\bar{a}}\ran\sim\omega^{-d/2},\qquad \lan \De k_{\bar{a}} \De k_{\bar{b}}\ran\sim\omega^{1-d/2},
\end{equation}
where the angle brackets denote the evaluation of the Gaussian integral, while every derivative $\partial/\partial k_{\bar{a}}$ entering into Taylor coefficients brings effectively the factor $\omega^{-1}$. It is easy to verify that the nontrivial leading contribution is of the order $\omega^{-1}$ in comparison with the naively expected estimate. Besides, in the leading order, $h_\al$ can be replaced by $h^{(0)}_\al$ and all the corrections related to the mass $m$ can be neglected as these are of the relative order $\omega^{-2}$.

Now we can estimate the behavior of $\bar{\Psi}_{\al\be}$ at the large momenta $|\spp_\al|\sim|\spp_\be|\sim\omega$. It follows from
\begin{equation}
    h^{(0)}_\al\sim(\omega V)^{-1/2},\qquad k_{\ddo}\sim\omega
\end{equation}
that
\begin{equation}\label{Psi_est}
    \bar{\Psi}_{\al\be}\sim\omega^{-1-d/2} V^{-1},
\end{equation}
where we have also retained the factors $V$. Then
\begin{equation}\label{part_num_est}
    n_\al=\sum_\be|\Psi_{\al\be}|^2=V\int\frac{d\spp_\be}{(2\pi)^d}|\Psi_{\al\be}|^2\sim\omega^{-2}V^{-1}.
\end{equation}
Of course, the estimate \eqref{Psi_est} is not valid for small momenta, but the main contribution to the integral \eqref{part_num_est} comes from the large momenta due to their large phase volume. Since, in a general position, the domain of $(\spp_\be,\spp_\al)$ admissible by \eqref{stat_p1}, \eqref{stat_p2} is open and conical, its volume grows as $\omega^{2d}$ for large $\omega$ and
\begin{equation}\label{part_num_tot}
    \sum_\al n_\al=V\int\frac{d\spp_\al}{(2\pi)^d} n_\al
\end{equation}
diverges as $\omega^{d-2}$ for $D\geqslant3$. For $d=2$ the number of created particles diverges logarithmically. It is not difficult to write down the explicit expression for the asymptotics of $\bar{\Psi}_{\al\be}$ at large momenta, but it is rather huge. In the next section, we shall find it under the assumption that $g_{\mu\nu}$ is close to $\eta_{\mu\nu}$. Notice that the curvature of the background metric cannot improve the convergence of \eqref{part_num_tot} in the ultraviolet domain. The only component of the above formulas that changes on a curved background is the commutator Green function. However, in the Riemann normal coordinates for the  spacetime metric, its leading asymptotics at large momenta coincides with the flat spacetime expression (see, e.g., \cite{DeWGAQFT.11}). Therefore, one reverts to the case considered above. This implies, in particular, that the inclusion of the effect of observers on the metric field does not change the asymptotics \eqref{part_num_est} and its consequences.

Strictly speaking, we ought to consider the ultraviolet behavior of $\Psi_{\al\be}$ constructed with the aid of $\tilde{G}^\La$ rather than $\tilde{G}$. However, for $\La$ much larger than all the momenta scales of the infinitely smooth background fields and the momenta of the states $\al$ and $\be$, the matrix element $\Psi_{\al\be}$ constructed by the use of $\tilde{G}^\La$ is close to the same matrix element associated with $\tilde{G}$, and in the limit $\La\rightarrow+\infty$ they coincide. Therefore, when the finite cutoff $\La$ is introduced, the asymptotics mentioned above is valid in the range
\begin{equation}\label{appl_range}
    \max(m,l^{-1})\ll |\spp_{\al,\be}|\ll\La,
\end{equation}
where $l$ is a characteristic scale of variations of the metric components measured by a congruence of observers.

To conclude this section, let us estimate the behavior of
\begin{equation}\label{Gab}
    \{\ups_\al,\ups_\be'\}=i\int d\spx\Big[\sqrt{|g'|}u_\al \big(g'^i\hat{P}'^\be_i+\frac{\omega_\be}{g'_{00}}\big)u'_\be -\sqrt{|g|}u'_\be \big(g^i\hat{P}^\al_i+\frac{\omega_\al}{g_{00}}\big)u_\al \Big]
\end{equation}
at large momenta $\spp_\al$, $\spp_\be$. The primed quantities on the right-hand side refer to the point $(y^0,\spx)$. The stationary points are found from the equations
\begin{equation}\label{stat_p3}
    p^\al_i+p'^\be_i=0.
\end{equation}
Repeating the above considerations almost word by word, we conclude that there is the map that assigns a pair $(\spp_\be,\spp_\al)$ to every pair $(\spx,p_i^\al)$. In a general position, the domain of $(\spp_\be,\spp_\al)$ corresponding to solutions of \eqref{stat_p3} is open and conical, the two terms in \eqref{Gab} do not cancel, and the integrand is not zero at the stationary points \eqref{stat_p3}. Therefore,
\begin{equation}
    \sum_{\al,\be}|\{\ups_\al,\ups_\be'\}|^2=V^2\int\frac{d\spp_\al d\spp_\be}{(2\pi)^{2d}}|\{\ups_\al,\ups_\be'\}|^2\sim \omega^d.
\end{equation}
Thus the Fock spaces constructed by the use of the creation-annihilation operators associated with the mode functions $\ups_\al$ and $\ups'_\al$ are not unitary equivalent, in a general position.

\section{Small perturbations of the metric components}\label{Pert_Anal}

Let us specialize the above considerations to the case when
\begin{equation}
    \vk^a(x)=\de^a_\mu x^\mu+\e\la^a(x),
\end{equation}
where $\la^a(x)$ are infinitely smooth functions with a compact support and $\e$ is a small parameter. Then, in the leading order in $\e$, we have
\begin{equation}
\begin{gathered}
    g^{ij}=-\de^{ij}-\e h^{ij},\qquad g_{ij}=-\de_{ij}+\e h^{ij},\qquad g_i=\e h_i=-g^i,\qquad g_{00}=1+\e h_{00},\\
    e^a_\mu=\de^a_\mu+\e\partial_\mu\la^a,\qquad e^\mu_a=\de^\mu_a-\e\partial_a\lambda^\mu,
\end{gathered}
\end{equation}
where
\begin{equation}
    h^{ij}=\partial^{(i}\lambda^{j)},\qquad h_i=\partial_i\lambda_0+\dot{\lambda}_i,\qquad h_{00}=2\dot{\la}_0,
\end{equation}
the indices are raised and lowered by the metric $\eta_{\mu\nu}$, and the dot denotes the derivative with respect to $x^0$. Also
\begin{equation}\label{metr_det}
    \sqrt{|g|}=1+\e\partial_\mu\la^\mu.
\end{equation}
The Hamilton equations \eqref{Ham_eqs} can be solved perturbatively up the first order in $\e$. The result is
\begin{equation}\label{pert_sol}
\begin{gathered}
    x_\perp^i=\s^i+\e\int_0^{t_0} d\tau\psi_\perp^i(x_\parallel(0)-\omega_\al\tau,\s),\qquad x_\parallel=x_\parallel(0) -\omega_\al t_0,\\
    p_i=p_i(0)+\e\int_0^{t_0} d\tau\partial_i\phi(x_\parallel(0)-\omega_\al\tau,\s),
\end{gathered}
\end{equation}
where
\begin{equation}
\begin{gathered}
    \phi(x)=\frac12 h_{00} +h_k n^\al_k +\frac12 h^{kl}n^\al_k n^\al_l,\qquad \psi^i(x)=-\omega_\al(h^{ij}n_j^\al +h_i),\qquad t_0=\omega_\al^{-1}(x_\parallel(0)-x_\parallel),\\
    x_\parallel^i=n_i^\al n_j^\al x^j=n_i^\al x_\parallel,\qquad x_\perp^i=x^i-n_i^\al x_\parallel,\\
    n_i^\al=p_i(0)/\omega_\al,\qquad \omega_\al=|\spp(0)|.
\end{gathered}
\end{equation}
In particular, it follows from \eqref{pert_sol} that the Jacobian appearing in \eqref{WKB_Jack} equals
\begin{equation}\label{WKB_Jack_app}
    \det(\partial x^i/\partial(\tau,\s))=1+O(\e).
\end{equation}
Substituting \eqref{pert_sol} into the Hamiltonian action with the Hamiltonian \eqref{Hamiltonian} and rescaling the integration variable, we arrive at
\begin{equation}\label{HJ_act}
    S_\al(x)=\omega_\al(x_\parallel-x_\parallel(0))+\e\omega_\al \int_0^{\bar{t}_0}d\tau\bar{\phi}(x_\parallel(0)-\tau,x_\perp),
\end{equation}
where
\begin{equation}
    \bar{t}_0=x_\parallel(0)-x_\parallel,\qquad\bar{\phi}(x)=\dot{\la}_0+(\partial_k\la_0+\dot{\la}_k)n_k^\al+\partial_k\la_l n_k^\al n_l^\al.
\end{equation}
The approximate mode functions are obtained by substituting \eqref{HJ_act}, \eqref{WKB_Jack_app}, \eqref{metr_det} into \eqref{WKB_form}, \eqref{WKB_norm}.

As is seen from \eqref{HJ_act}, these mode functions cannot be obtained at any finite order of the standard Rayleigh-Schr\"{o}dinger perturbation theory with respect to $\e$. Physically, this is related to the fact that the strength of the gravitational interaction grows linearly with energy on a classical level due to the equivalence principle. The higher the energy of a particle (real or virtual), the stronger its interaction with the gravitational field. Therefore, one cannot use the usual perturbation theory in the ultraviolet limit. Mathematically, non-perturbativity in the ultraviolet domain stems from the fact that the metric field enters into the principal symbol of the wave operator governing the dynamics of a scalar field. In contrast to gravity, the strength of the electromagnetic interaction, for example, does not depend on the energy on a classical level. Therefore, in the ultraviolet limit, the acceleration experienced by a particle goes to zero, and the perturbation theory may be applied to obtain the mode functions. This also leads to better behavior of $\Psi_{\al\be}$ in the ultraviolet domain such that $\Psi_{\al\be}$ is Hilbert-Schmidt in QED with respect to a uniform congruence of inertial observers in the Minkowski spacetime \cite{Scharf79}.

The stationary condition \eqref{stat_p1} for the approximate mode functions looks as
\begin{equation}\label{stat_p1_app}
\begin{split}
    p_i^\al(0)+q_i+\e\Big[\omega_\al\int_0^{\bar{t}_0}d\tau \frac{\partial\bar{\phi}(x_\parallel(0)-\tau,x_\perp)}{\partial x_\perp^i} -\omega_\al n_i^\al\bar{\phi}(x) +q_a\partial_i\la^a\Big]&=0,\\
    p_i^\be(0)-q_i+\e\Big[\omega_\be\int_0^{\bar{t}'_0}d\tau \frac{\partial\bar{\phi}'(y_\parallel(0)-\tau,y_\perp)}{\partial y_\perp^i} -\omega_\be n_i^\be\bar{\phi}'(y) -q_a\partial_i\la'^a\Big]&=0,\\
    n_{\bar{a}}(x^0-y^0)+x^{\bar{a}}-y^{\bar{a}} +\e\Big[n_{\bar{a}}(\la^0-\la'^0)+\la^{\bar{a}}-\la'^{\bar{a}} \Big]&=0,
\end{split}
\end{equation}
where the primes remind us that the respective quantities refer to the point $y$. These equations ought to be solved with respect to $x^{\bar{a}}$, $y^{\bar{a}}$, and $q_{\bar{a}}$ (the indices $\bar{a}$ and $i$ can be identified) for given $p_i^\al(0)$ and $p_i^\be(0)$. Setting
\begin{equation}
    p_i^\be(0)=:-p_i^\al(0) +\e\De_i,
\end{equation}
and introducing the notation
\begin{equation}
    q_i=q_i^{(0)}+\e q_i^{(1)}+\cdots,\qquad x^i=x^i_{(0)}+\e x^i_{(1)}+\cdots,\qquad y^i=y^i_{(0)}+\e y^i_{(1)}+\cdots,
\end{equation}
we obtain in the leading order
\begin{equation}\label{zero_ord}
    q_i^{(0)}=-p_i^\al(0),\qquad x^i_{(0)}=y^i_{(0)}+n^\al_i(x^0-y^0).
\end{equation}
The next order equations are written as
\begin{equation}\label{first_ord}
\begin{split}
    q_i^{(1)} +q^{(0)}_a\partial_i\lambda^a(x_{(0)})+\omega_\al \int_0^{\bar{t}_0}d\tau \frac{\partial\bar{\phi}(x_\parallel(0)-\tau,x_\perp)}{\partial x_\perp^i}\Big|_{x= x_{(0)}} -\omega_\al n_i^\al\bar{\phi}(x_{(0)})&=0,\\
    q_i^{(1)} +q^{(0)}_a\partial_i\lambda'^a(y_{(0)})-\omega_\al \int_0^{\bar{t}'_0}d\tau \frac{\partial\bar{\phi}'(y_\parallel(0)-\tau,y_\perp)}{\partial y_\perp^i}\Big|_{y= y_{(0)}} -\omega_\al n_i^\al\bar{\phi}'(y_{(0)})&=\De_i,\\
    n_{\bar{a}}^{(1)}(x^0-y^0) +x^{\bar{a}}_{(1)}-y^{\bar{a}}_{(1)} -n^\al_{\bar{a}}\big[\lambda^0(x_{(0)})-\lambda'^0(y_{(0)})\big] +\la^{\bar{a}}(x_{(0)}) -\la'^{\bar{a}}(y_{(0)})&=0.
\end{split}
\end{equation}
The above equations can be solved as follows. For fixed $y^i_{(0)}$, one takes $q_i^{(1)}$ from the second equation in \eqref{first_ord} and substitutes it into the first equation, where $x^i_{(0)}$ is expressed through $y^i_{(0)}$ by means of the second equation in \eqref{zero_ord}. In a general position, the resulting equation can be resolved with respect to $y^i_{(0)}$. Substituting this solution into the second equation in \eqref{first_ord}, we obtain $q_i^{(1)}$ and so $n_{\bar{a}}^{(1)}$. Then the third equation in \eqref{first_ord} gives the relation between $x^i_{(1)}$ and $y^i_{(1)}$ that should be used in the next step of the perturbation theory.

Introducing the notation
\begin{equation}
    \Si_1(x,y,k):=S_\al(x)+S_\be(y)+k_a[\vk^a(x)-\vk^a(y)],
\end{equation}
we find
\begin{equation}\label{ABC}
\begin{gathered}
    \frac{\partial\Si_1}{\partial x^i\partial k_{\bar{a}}}\equiv B^{\bar{a}}_i=\de^{\bar{a}}_i+O(\e),\qquad \frac{\partial\Si_1}{\partial y^i\partial k_{\bar{a}}}\equiv B'^{\bar{a}}_i=-\de^{\bar{a}}_i+O(\e),\\
    \frac{\partial\Si_1}{\partial k_{\bar{a}}\partial k_{\bar{b}}}\equiv C^{\bar{a}\bar{b}}=\frac{\pr_{\bar{a}\bar{b}}}{q_{\ddo}}(x^0-y^0)+O(\e),
\end{gathered}
\end{equation}
and
\begin{equation}\label{ABCa}
\begin{split}
    \frac{\partial\Si_1}{\partial x^i\partial x^j}\equiv A_{ij}=\,&\e\Big[q_a\partial_{ij}\la^a -\omega_\al n_i^\al n_j^\al \frac{\partial\bar{\phi}(x)}{\partial x_\parallel} \\
    &-\omega_\al n^\al_{(i}\frac{\partial\bar{\phi}(x)}{\partial x_\perp^{j)}} +\omega_\al\int_0^{\bar{t}_0}d\tau\frac{\partial\bar{\phi}(x_\parallel(0)-\tau,x_\perp)}{\partial x_\perp^i \partial x_\perp^j} \Big]+O(\e^2),\\
    \frac{\partial\Si_1}{\partial y^i\partial y^j}\equiv A'_{ij}=\,&\e\Big[-q_a\partial_{ij}\la'^a -\omega_\al n_i^\al n_j^\al \frac{\partial\bar{\phi}'(y)}{\partial y_\parallel} \\
    &+\omega_\al n^\al_{(i}\frac{\partial\bar{\phi}'(y)}{\partial y_\perp^{j)}} +\omega_\al\int_0^{\bar{t}'_0}d\tau \frac{\partial\bar{\phi}'(y_\parallel(0)-\tau,y_\perp)}{\partial y_\perp^i \partial y_\perp^j} \Big]+O(\e^2),
\end{split}
\end{equation}
where $\pr_{\bar{a}\bar{b}}=\de_{\bar{a}\bar{b}}-n_{\bar{a}}n_{\bar{b}}$, the derivatives are taken at the stationary point \eqref{stat_p1_app}, and only the leading in $\omega$ terms are retained. Besides, in this leading order,
\begin{equation}
    \Si_1(x,y,k)|_{\text{st.p.}}=S_\al(x)|_{\text{st.p.}}+S_\be(y)|_{\text{st.p.}}.
\end{equation}
The quadratic form of the Gaussian integral is given by
\begin{equation}\label{quadr_form}
    G^{-1}:=\left[
       \begin{array}{ccc}
         A & 0 & B \\
         0 & A' & B' \\
         B^T & B'^T & C \\
       \end{array}
     \right].
\end{equation}
Employing a blockwise inversion formula, it is not difficult to obtain the inverse of \eqref{quadr_form},
\begin{equation}
    G=\left[
         \begin{array}{ccc}
           (A+A')^{-1} & (A+A')^{-1} & (A+A')^{-1}A' \\
           (A+A')^{-1} & (A+A')^{-1} & -(1+A^{-1}A')^{-1} \\
           A'(A+A')^{-1} & -(1+A'A^{-1})^{-1} & -A'(A+A')^{-1}A \\
         \end{array}
       \right],
\end{equation}
and the determinant
\begin{equation}
    \det G^{-1}=(-1)^d\det(A+A'),
\end{equation}
in the leading order in $\e$. Notice that
\begin{equation}
    A\sim A'\sim \e\omega.
\end{equation}
In order to obtain the contribution of the first term in the curly brackets in \eqref{Psi_bar_WKB} in the leading order in $\e$ and $\omega$, we also need the expansion
\begin{equation}
    \vf(y):=k'^0-g'^i P'^\be_i -\frac{\omega_\be}{g'_{00}}\approx q_{\ddo}-\omega_\al-\e\big[q^b\partial_b\lambda'_0 +\omega_\al n^\al_i (\dot{\la}'_i+\partial_i\la_0') -2\omega_\al\dot{\la}'_0-n_i^\al\De_i \big]=:\vf'_0+\e\vf'_1.
\end{equation}
Hence,
\begin{equation}\label{vf_der}
\begin{gathered}
    \frac{\partial\vf_0'}{\partial k_{\bar{a}}}=-n^\al_{\bar{a}},\qquad \frac{\partial\vf_1'}{\partial y^i}=\omega_\al(\partial_i\dot{\la}'_0 -n^\al_j\partial_i\dot{\la}'_j -2n^\al_j\partial_{ij}\la'_0),\\ \frac{\partial\vf_1'}{\partial y^i \partial y^j}=\omega_\al(\partial_{ij}\dot{\la}'_0 -n^\al_k\partial_{ij}\dot{\la}'_k -2n^\al_k\partial_{ijk}\la'_0),
\end{gathered}
\end{equation}
at the stationary point \eqref{stat_p1_app}. The leading order contribution comes from the Gaussian integrals that schematically can be represented in the form
\begin{equation}\label{Gauss_ints}
\begin{split}
    (a1)_1:=&\,\frac{i}{6}\frac{\partial\vf_0'}{\partial k_{\bar{a}}}\frac{\partial\Si_1}{\partial y^i \partial y^j \partial y^k} \lan\De k_{\bar{a}}\De y^i\ran \lan\De y^j\De y^k\ran,\\
    (a2)_1:=&\,\frac{i}{6}\frac{\partial\vf_0'}{\partial k_{\bar{a}}}\frac{\partial\Si_1}{\partial x^i \partial x^j \partial x^k} \lan\De k_{\bar{a}}\De x^i\ran \lan\De x^j\De x^k\ran,\\
    (b1)_1:=&\,\frac{i}{6}\e \frac{\partial\vf_1'}{\partial y^i}\frac{\partial\Si_1}{\partial y^j \partial y^k \partial y^l} \lan \De y^i \De y^j\ran \lan\De y^k\De y^l\ran,\\
    (b2)_1:=&\,\frac{i}{6}\e \frac{\partial\vf_1'}{\partial y^i}\frac{\partial\Si_1}{\partial x^j \partial x^k \partial x^l} \lan \De y^i \De x^j\ran \lan\De x^k\De x^l\ran,\\
    (c)_1:=&\,\frac{\e}{2} \frac{\partial\vf_1'}{\partial y^i \partial y^j} \lan \De y^i \De y^j\ran.
\end{split}
\end{equation}
These contributions are all of the order $\e^0\omega^{-1}$ with respect to the naively expected estimate of $\bar{\Psi}_{\al\be}$. The third derivative of $\Si_1$ is obtained easily from \eqref{ABCa}.

The generating function for evaluation of the Gaussian type integrals is written in our case as
\begin{equation}
    Z(J)=\int\frac{d\spx d\spk d\spy}{(2\pi)^d}\exp\Big\{\frac{i}{2} \left[
                                                           \begin{array}{ccc}
                                                             \spx & \spy & \spk \\
                                                           \end{array}
                                                         \right]
     G^{-1}
     \left[
       \begin{array}{c}
         \spx \\
         \spy \\
         \spk \\
       \end{array}
     \right]
     +i J\left[\begin{array}{ccc}
               \spx & \spy & \spk \\
               \end{array}
               \right]\Big\}=\frac{(2\pi i)^{d/2}}{\det^{1/2}(A+A')}e^{-iJ^TGJ/2},
\end{equation}
in the leading order. Here the principal branch of the square root is taken. The common factor at the Gaussian integrals becomes
\begin{equation}
    f^1_{\al\be}=\frac{\sqrt{gg'}}{2\omega_\al V}\frac{(2\pi i)^{d/2}}{\det^{1/2}(A+A')}e^{iS_\al+iS'_\be}.
\end{equation}
Thus,
\begin{equation}
\begin{split}
    (a1)_1=&\, -\frac{i}{2}\frac{\partial\Si_1}{\partial y^i \partial y^j \partial y^k}n^\al_{\bar{a}}[A(A+A')^{-1}]_{\bar{a}i} (A+A')^{-1}_{jk},\\
    (a2)_1=&\, \frac{i}{2} \frac{\partial\Si_1}{\partial x^i \partial x^j \partial x^k} n^\al_{\bar{a}}[A'(A+A')^{-1}]_{\bar{a}i} (A+A')^{-1}_{jk},\\
    (b1)_1=&\, -\frac{i}{2}\e \frac{\partial\vf_1'}{\partial y^i} \frac{\partial\Si_1}{\partial y^j \partial y^k \partial y^l} (A+A')^{-1}_{ij}(A+A')^{-1}_{kl},\\
    (b2)_1=&\, -\frac{i}{2} \e \frac{\partial\vf_1'}{\partial y^i} \frac{\partial\Si_1}{\partial x^j \partial x^k \partial x^l} (A+A')^{-1}_{ij}(A+A')^{-1}_{kl},\\
    (c)_1=&\, \frac{i}{2}\e \frac{\partial\vf_1'}{\partial y^i \partial y^j} (A+A')^{-1}_{ij},
\end{split}
\end{equation}
and all the contributions must be multiplied by the factor $f^1_{\al\be}$. The dependence on $\De_i$ enters into these terms only through the position of a stationary point. Substituting the explicit expression for $A'$ into $(a2)_1$, one can verify that the contributions $(a2)_1$ and $(b2)_1$ cancel out.

The analogous analysis applies to the second term in the curly brackets in \eqref{Psi_bar_WKB}. The corresponding formulas are almost the same as above. The approximate stationarity condition reads as
\begin{equation}\label{stat_p2_app}
\begin{split}
    p_i^\al(0)-q_i+\e\Big[\omega_\al\int_0^{\bar{t}_0}d\tau \frac{\partial\bar{\phi}(x_\parallel(0)-\tau,x_\perp)}{\partial x_\perp^i} -\omega_\al n_i^\al\bar{\phi}(x) -q_a\partial_i\la^a\Big]&=0,\\
    p_i^\be(0)+q_i+\e\Big[\omega_\be\int_0^{\bar{t}'_0}d\tau \frac{\partial\bar{\phi}'(y_\parallel(0)-\tau,y_\perp)}{\partial y_\perp^i} -\omega_\be n_i^\be\bar{\phi}'(y) +q_a\partial_i\la'^a\Big]&=0,\\
    n_{\bar{a}}(x^0-y^0)+x^{\bar{a}}-y^{\bar{a}} +\e\Big[n_{\bar{a}}(\la^0-\la'^0)+\la^{\bar{a}}-\la'^{\bar{a}} \Big]&=0.
\end{split}
\end{equation}
Introducing the function
\begin{equation}
    \Si_2(x,y,k):=S_\al(x)+S_\be(y)-k_a[\vk^a(x)-\vk^a(y)],
\end{equation}
we obtain in the leading order
\begin{equation}\label{ABC2}
\begin{gathered}
    \frac{\partial\Si_2}{\partial x^i\partial k_{\bar{a}}}\equiv B^{\bar{a}}_i=-\de^{\bar{a}}_i+O(\e),\qquad \frac{\partial\Si_2}{\partial y^i\partial k_{\bar{a}}}\equiv B'^{\bar{a}}_i=\de^{\bar{a}}_i+O(\e),\\
    \frac{\partial\Si_2}{\partial k_{\bar{a}}\partial k_{\bar{b}}}\equiv C^{\bar{a}\bar{b}}=\frac{\pr_{\bar{a}\bar{b}}}{q_{\ddo}}(y^0-x^0)+O(\e),
\end{gathered}
\end{equation}
and
\begin{equation}
\begin{split}
    \frac{\partial\Si_2}{\partial x^i\partial x^j}\equiv A_{ij}=&\,\e\Big[-q_a\partial_{ij}\la^a -\omega_\al n_i^\al n_j^\al \frac{\partial\bar{\phi}(x)}{\partial x_\parallel}\\
    &-\omega_\al n^\al_{(i}\frac{\partial\bar{\phi}(x)}{\partial x_\perp^{j)}} +\omega_\al\int_0^{\bar{t}_0}d\tau\frac{\partial\bar{\phi}(x_\parallel(0)-\tau,x_\perp)}{\partial x_\perp^i \partial x_\perp^j} \Big]+O(\e^2),\\
    \frac{\partial\Si_2}{\partial y^i\partial y^j}\equiv A'_{ij}=&\,\e\Big[q_a\partial_{ij}\la'^a -\omega_\al n_i^\al n_j^\al \frac{\partial\bar{\phi}'(y)}{\partial y_\parallel}\\
    &+\omega_\al n^\al_{(i}\frac{\partial\bar{\phi}'(y)}{\partial y_\perp^{j)}} +\omega_\al\int_0^{\bar{t}'_0}d\tau \frac{\partial\bar{\phi}'(y_\parallel(0)-\tau,y_\perp)}{\partial y_\perp^i \partial y_\perp^j} \Big]+O(\e^2),
\end{split}
\end{equation}
at the solutions to \eqref{stat_p2_app}. Of course, $A$, $A'$, $B$, $B'$, and $C$ appearing in this formula do not coincide, in general, with the similar quantities in formulas \eqref{ABC}, \eqref{ABCa}. The expression standing in the exponent becomes
\begin{equation}
    \Si_2(x,y,k)|_{\text{st.p.}}=S_\al(x)|_{\text{st.p.}}+S_\be(y)|_{\text{st.p.}},
\end{equation}
at the stationary points in the leading order in $\omega$. The quadratic form of the Gaussian integral is written as \eqref{quadr_form}. Its inverse and the determinant are
\begin{equation}
    G=\left[
         \begin{array}{ccc}
           (A+A')^{-1} & (A+A')^{-1} & -(A+A')^{-1}A' \\
           (A+A')^{-1} & (A+A')^{-1} & (1+A^{-1}A')^{-1} \\
           -A'(A+A')^{-1} & (1+A'A^{-1})^{-1} & -A'(A+A')^{-1}A \\
         \end{array}
       \right],\qquad \det G^{-1}=(-1)^d\det(A+A'),
\end{equation}
in the leading order. We also need the expansion
\begin{equation}
    \psi(x):=k^0-g^i P^\al_i -\frac{\omega_\al}{g_{00}}\approx q_{\ddo}-\omega_\al-\e\big[q^b\partial_b\lambda_0 -\omega_\al n^\al_i (\dot{\la}_i+\partial_i\la_0) -2\omega_\al\dot{\la}_0\big]=:\psi_0+\e\psi_1,
\end{equation}
whence
\begin{equation}\label{psi_der}
\begin{gathered}
    \frac{\partial\psi_0}{\partial k_{\bar{a}}}=n^\al_{\bar{a}},\qquad \frac{\partial\psi_1}{\partial x^i}=\omega_\al(\partial_i\dot{\la}_0 +n^\al_j\partial_i\dot{\la}_j +2n^\al_j\partial_{ij}\la_0),\\
    \frac{\partial\psi_1}{\partial x^i \partial x^j}=\omega_\al(\partial_{ij}\dot{\la}_0 +n^\al_k\partial_{ij}\dot{\la}_k +2n^\al_k\partial_{ijk}\la_0),
\end{gathered}
\end{equation}
at the stationary points \eqref{stat_p2_app}. The Gaussian integrals giving the leading order contribution to $\bar{\Psi}_{\al\be}$ have the same form as \eqref{Gauss_ints} with the replacement of the derivatives \eqref{vf_der} by \eqref{psi_der} and $\Si_1$ by $\Si_2$. We shall distinguish these integrals by the index $2$. The common factor at the Gaussian integrals is given by
\begin{equation}
    f^2_{\al\be}=-\frac{\sqrt{gg'}}{2\omega_\al V}\frac{(2\pi i)^{d/2}}{\det^{1/2}(A+A')}e^{iS_\al+iS'_\be}.
\end{equation}
Thus, evaluating the Gaussian integrals, we have
\begin{equation}
\begin{split}
    (a1)_2=&\, -\frac{i}{2}\frac{\partial\Si_2}{\partial y^i \partial y^j \partial y^k}n^\al_{\bar{a}}[A(A+A')^{-1}]_{\bar{a}i} (A+A')^{-1}_{jk},\\
    (a2)_2=&\, \frac{i}{2} \frac{\partial\Si_2}{\partial x^i \partial x^j \partial x^k} n^\al_{\bar{a}}[A'(A+A')^{-1}]_{\bar{a}i} (A+A')^{-1}_{jk},\\
    (b1)_2=&\, -\frac{i}{2}\e \frac{\partial\psi_1}{\partial x^i} \frac{\partial\Si_2}{\partial y^j \partial y^k \partial y^l} (A+A')^{-1}_{ij}(A+A')^{-1}_{kl},\\
    (b2)_2=&\, -\frac{i}{2} \e \frac{\partial\psi_1}{\partial x^i} \frac{\partial\Si_2}{\partial x^j \partial x^k \partial x^l} (A+A')^{-1}_{ij}(A+A')^{-1}_{kl},\\
    (c)_2=&\, \frac{i}{2}\e \frac{\partial\psi_1}{\partial x^i \partial x^j} (A+A')^{-1}_{ij},
\end{split}
\end{equation}
up to the common factor $f^2_{\al\be}$. The contributions $(a1)_2$ and $(b1)_2$ cancel out.

Collecting all the contributions together, we obtain
\begin{equation}\label{Psi_asymp}
    \bar{\Psi}_{\al\be}\approx (a1)_1+(b1)_1+(c)_1+(a2)_2+(b2)_2+(c)_2\sim \omega^{-1-d/2}V^{-1}\e^{-d/2}.
\end{equation}
The structure of $\bar{\Psi}_{\al\be}$ is of the form
\begin{equation}
    a e^{i\Si_1}+b e^{i\Si_2}.
\end{equation}
As long as
\begin{equation}
    |a e^{i\Si_1}+b e^{i\Si_2}|\geqslant||a|-|b||,
\end{equation}
and $|a|=|b|$ on the set of measure zero, the oscillating factors $e^{i\Si_1}$, $e^{i\Si_1}$ cannot improve the convergence of the integral \eqref{part_num_tot} in the ultraviolet domain. Therefore, we conclude, just as in the previous section, that \eqref{part_num_tot} diverges for $D\geqslant3$. It may happen that $\det(A+A')=0$ for certain momenta $\spp_\al$, $\spp_\be$. Then the formulas above are not valid, and more sophisticated analysis should be used to obtain the ultraviolet asymptotics (see, e.g., \cite{Fedoryuk.6}). Nevertheless, this occurs on the set of points of measure zero in the space of $(\spp_\al,\spp_\be)$ and so it does not affect our results.

As we have already discussed in the previous section, the expression \eqref{Psi_asymp} can be used to describe the ultraviolet asymptotics of \eqref{Psi_def} for finite $\La$. Since this expression was obtained by the use of the commutator Green function in the regularization removal limit, the asymptotics \eqref{Psi_asymp} is valid only in the region of momenta \eqref{appl_range}. It is not difficult to find a loose estimate for the average number of produced particles. To this end, it is convenient to perform a change of variables in the integral \eqref{part_num_est} and replace $\spp_\be$ by the stationary points $\spy$ obtained from \eqref{stat_p1_app}, \eqref{stat_p2_app}. The respective Jacobian
\begin{equation}
    \det\frac{\partial p^\be_i(0)}{\partial y^j}\sim (\e\omega)^d.
\end{equation}
The domain of nondegenerate stationary points $\spy$ lies in the region, which is a union of the region with $g_{\mu\nu}(y)\neq\eta_{\mu\nu}$ and the domain obtained from the region $g_{\mu\nu}(x)\neq\eta_{\mu\nu}$ by emanating null geodesics to the past up to their intersection with the hypersurface related to the state $\be$. Let us denote by $\Omega$ the volume of this region. Then the average number of particles in the state $\al$ is
\begin{equation}\label{part_num_re}
    n_\al\sim\frac{\Omega}{V\omega^2 l^2}.
\end{equation}
Notice that the number of created particles is independent of $\e$. The factor $V^{-1}$ comes from the normalization of states to unity and disappears when multiplied by the density of states in the momentum space.

\section{Conclusion}

Let us summarize the results. We obtained the explicit expression for the ultraviolet asymptotics of the number of particles created from the vacuum in the Minkowski spacetime with respect to a congruence of observers of a general form. The splitting into positive- and negative-frequency modes was defined with aid of diagonalization of the instantaneous Hamiltonian of quantum fields. We found that, in the regularization removal limit, the total number of created particles diverges in the ultraviolet domain for a general congruence of observers in the $D$-dimensional spacetime with $D\geqslant3$. The inclusion of the metric curvature does not improve the convergence. That conclusion is in agreement with the similar calculations made for cosmological metric backgrounds with a special choice of the congruence of observers \cite{Park1,FullingAQFT,GriMaMos.11,ParkTom}. The same conclusion but for a different splitting of the modes into positive- and negative-frequency ones was drawn in \cite{TorVara}. This is a characteristic feature of the gravitational interaction and it is absent, for example, for the background electromagnetic fields \cite{Scharf79}.

It is not clear at the present stage of research whether and how those particles created in the ultraviolet spectral range can be observed. In order to obtain the quantities observable in experiment, one should also take into account the vacuum polarization effects. This is important, for example, for evaluation of the average of the energy-momentum tensor. Besides, one should bear in mind that, due to universality of gravitational interaction, the mode functions of all the particles, including those the detector consists of, change accordingly with the different choices of a congruence of observers. One of the manifestations of this particle creation is the response of the Unruh detector \cite{Unruh,DeWGAQFT.11}, but it does not provide a complete picture, for it is described usually as a mathematical point. One should bear in mind that, in general, these created particles are not accumulated during the evolution and can be absorbed by the vacuum as well. As for common scattering processes on such backgrounds with energies much smaller than the cutoff scale, the only modification concerning the production of particles from the vacuum is the replacement of the ordinary probabilities by the inclusive ones similarly to the case of theories with infrared divergencies.

Nevertheless, on the formal level, the result that one cannot take a regularization removal limit for a general congruence of observers is rather spectacular. It implies, in particular, that, for generally defined $in$ and $out$ vacuum states, the imaginary part of the one-loop $in$-$out$ effective action diverges in the regularization removal limit. The explicit expression for this divergence readily follows from \eqref{Psi_asymp}. This divergence is ultraviolet, non-local in time, and imaginary. According to the general rules of renormalization theory (see, e.g., \cite{Collins.12,BogolShir}), it cannot be canceled out by the counterterms. In fact, the nonlocal in time prescriptions for the splitting of the mode functions into positive- and negative-frequency ones discussed in Introduction are equivalent to an addition of nonlocal in time counterterms to the Hamiltonian. If one does not add such counterterms, then, first, in considering quantum dynamics with respect to a general congruence of observers, one cannot take the regularization removal limit in advance, although some observables may prove to be finite when the cutoff is removed. In particular, the Heisenberg equation for a scalar quantum field \eqref{Heis_eqs} cannot bluntly be replaced by the Klein-Gordon equation, which emerges only in the regularization removal limit. Second, the dependence of QFT on a choice of a congruence of observers that was discussed, e.g., in \cite{gmse.11,KalKaz1,KalKaz2.12,qgadm,KazMil1,Isham.12,KucharPoT,KucharI,KucharII,Fulling} becomes evident, since the cutoff cannot be removed consistently. Even for static metrics with static congruences of observers in the $in$ and $out$ states, when the imaginary part of the one-loop effective action is well defined \cite{Dimock79,Wald79,FulNarWal}, one needs to know the intermediate quantum evolution in order to find, for example, the average of the energy-momentum tensor at a given point $x$. Having chosen the congruence of observers, this quantity becomes observable in the sense of general relativity. In describing the evolution at intermediate times, one inevitably encounters with the above unitarity issues, and so one cannot take beforehand the regularization removal limit in this case too. Not to mention the fact that a static metric with a static congruence of observers is an ideal situation which is never realized.

Of course, then the natural question arises. If quantum dynamics depend severely on the choice of a congruence of observers and, in the regularization removal limit, different choices of congruences result in unitary inequivalent quantum theories, then what is the ``correct'' choice of a congruence of observers leading to what we observe in experiments? In order to maintain general covariance, the congruence of observers and, in particular, the timelike vector field $\xi^\mu$ must possess their our dynamics. It was shown in \cite{gmse.11} that under rather general assumptions this vector field obeys the equations of motion of a relativistic fluid with a certain equation of state. This gives the answer to the above question, but does not provide clues to the nature of such a field. Whether this field is fundamental or composite? What is the exact equation of state of this fluid and how to describe consistently its quantum dynamics? Whether and how it interacts with the ordinary fields? Some of these questions were addressed in \cite{qgadm}, but they remain a subject for further research.

\paragraph{Acknowledgments.}

The work is supported by the Ministry of Education and Science of the Russian Federation, Project No. 3.9594.2017/8.9.

\end{document}